\documentclass[11pt, A4]{article}

\usepackage{algorithm}
\usepackage{algpseudocode}
\usepackage{ragged2e}
\usepackage{graphicx}
\usepackage{amssymb}
\usepackage{amsthm}
\usepackage{subfigure}
\usepackage{url}
\usepackage{amsmath}
\usepackage{color,soul} 
\usepackage[square,sort,comma,numbers]{natbib}
\usepackage[english]{babel}
\usepackage{authblk}
\usepackage{anysize}
\usepackage{float}
\usepackage{xcolor}

\newtheorem{finding}{Finding}

\marginsize{2cm}{2cm}{.3cm}{2cm}

\graphicspath{{figs/}}

\begin{document}

\title{On complexity of branching droplets in electrical field}
\author[1]{Mohammad Mahdi Dehshibi}
\author[2]{Jitka \v{C}ejkov\'{a} \thanks{Contact: jitka.cejkova@vscht.cz}}
\author[2]{Dominik \v{S}v\'{a}ra}
\author[3]{Andrew Adamatzky}

\affil[1]{Department of Computer Science, Universitat Oberta de Catalunya, Barcelona, Spain}
\affil[2]{University of Chemistry and Technology Prague,  Czech Republic}
\affil[3]{Unconventinal Computing Laboratory, University of the West England, Bristol, UK}

\maketitle

\begin{abstract}
Decanol droplets in a thin layer of sodium decanoate with sodium chloride exhibit bifurcation branching growth due to interplay between osmotic pressure, diffusion and surface tension. We aimed to evaluate if morphology of the branching droplets changes when the droplets are subject to electrical potential difference.  We analysed graph-theoretic structure of the droplets and applied several complexity measures. We found that, in overall, the current increases complexity of the branching droplets in terms of number of connected components and nodes in their graph presentations, morphological complexity and compressibility. 

\vspace{5mm}


\end{abstract}

\vspace{0.5cm}

\vspace{0.5cm}

\section {Introduction}

Organic droplets placed in a thin layer of surfactants solution containing alkali salts or hydroxides develop irregular shapes. In~\cite{vcejkova2018multi,cejkova2016evaporation} we reported results of a parametric study of  decanol droplets floating on aqueous decanoate solution with sodium chloride. We observed growth of branching structures. We found that, depending on the molar ratio of decanol, decanoate, and the added sodium chloride, three regimes  of behavior can be observed: 
(i)~the disintegration of a decanol droplet into smaller droplets, 
(ii) tentacle-like pattern formation, and 
(iii) no response from the droplet.  The switch between the above morphological regimes is governed by the amount of sodium chloride in the solution. 
In present paper we advance our previous ideas into the field of electrical field based morphogenesis. We subject decanol/deconate system~\cite{cejkova2016evaporation} to electrical field and analyse the branching structures formed. To quantify differences between control branching droplets and the droplets subjected to an electrical field we use graph-theoretical indicators (numbers of disconnected components, nodes and leaves) and Lempel-Ziv (LZ) complexity~\cite{lempel1976complexity}) of the snapshots of the droplets. 

\section{Methods}

Chemicals used were sodium decanoate (Sigma), sodium chloride (Penta), sodium hydroxide (Penta) and decanol (Fluka). All chemicals were used as obtained without any further purification. Deionised water was produced using an ionex filter (Aqual 25, Czech Republic). Sodium decanoate solution was prepared as 10~mM by dissolving of sodium decanoate in water and adjusting the pH to 12-13 (by using 5~M sodium hydroxide solution). The sodium chloride solution in water was prepared as saturated (i.e., concentration 6.5~M). Oil Red O (Sigma) was used to colour of decanol droplets for better visualisation (concentration ca 2~mg/ml).

The cell used for experiments was prepared as follows. The microscopic glass slide ($76 \times 26~mm$) was partly covered with the adhesive tape to block the parts of the slide that need to stay nonconductive. The slide was then inserted into the sputtering device (sputter coater Emitech K550X) where uncovered parts were covered with a thin gold layer. The tape was removed from the glass when the process of the coating was finished. The final step of the process was the application of the hydrophobic coating. The hydrophobic coating was essential for keeping the aqueous decanoate droplet in a defined space between electrodes. The microscopic cover slides were used as a mask in this process. As two shapes of the glass slides were used, two versions of the device were created. The square shape cover slide ($18 \times 18~mm$) was used in one version, and the round shaped slide ($18~mm$ diameter) was used in another version.

The experiments were performed as follows. 250~$\mu$l of 10~mM sodium decanoate solution was spread on the round or  square shaped non-hydrophobized part of the glass slides. The decanol droplet (2.4~$\mu$l) was then placed on the layer of the decanoate. The saturated solution of sodium chloride (1.5~$\mu$l) was added to the decanoate layer. The slide was connected to the power source  through golden electrodes. Experiments were done  for voltages in the range of 0.5-3.5~V, the power supply used for experiments was Korad KA3305P. All experiments were repeated five times, and the experiments were captured in time  with the camera (DFK 23U274 Imaging Source) from the top view. 
All versions of the experimental devices were tested without the application of the electric field to confirm that the hydrophobic coating and the presence of golden electrodes do not affect the droplet behaviour.

\begin{figure}[!tbp]
    \centering
    \subfigure[]{\includegraphics[scale=0.45]{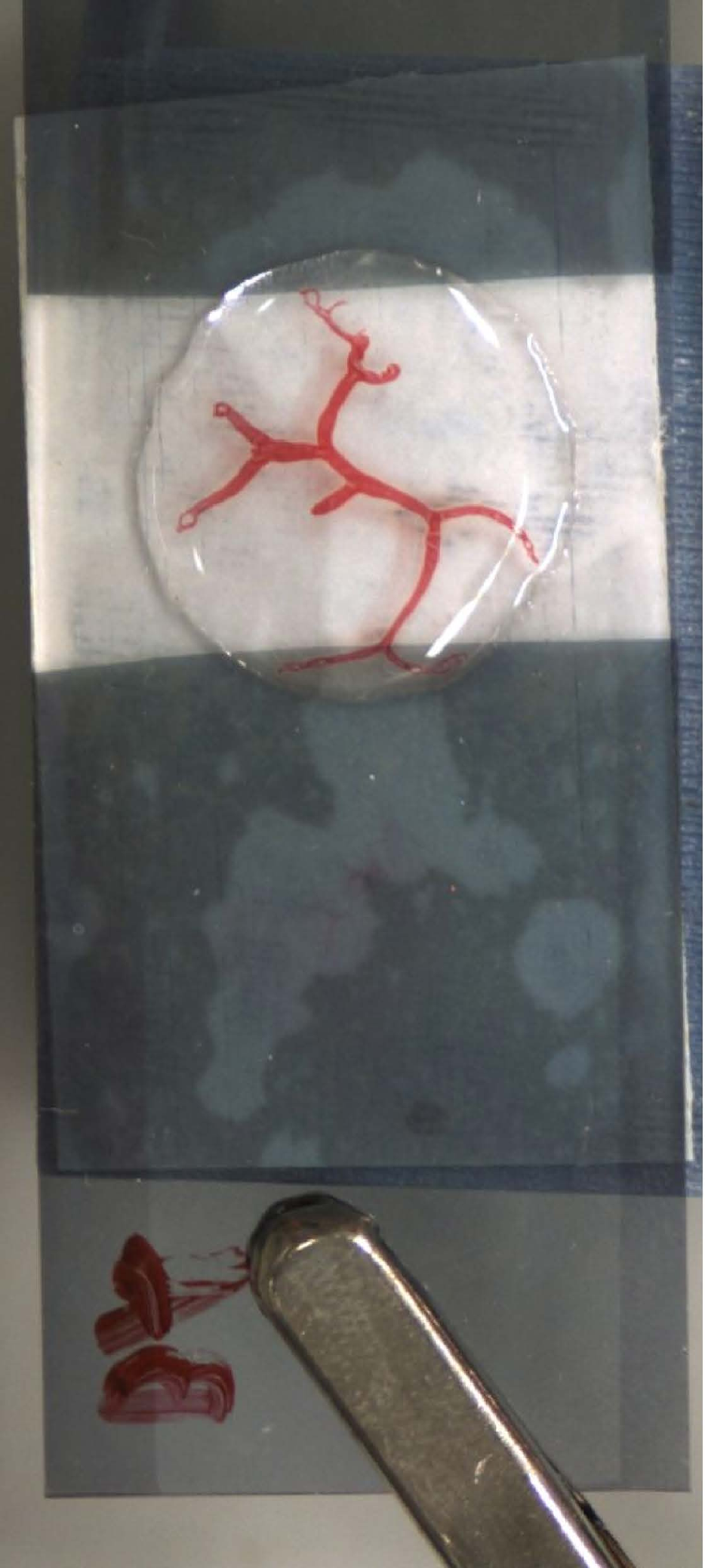}}
    \subfigure[]{\includegraphics[scale=0.45]{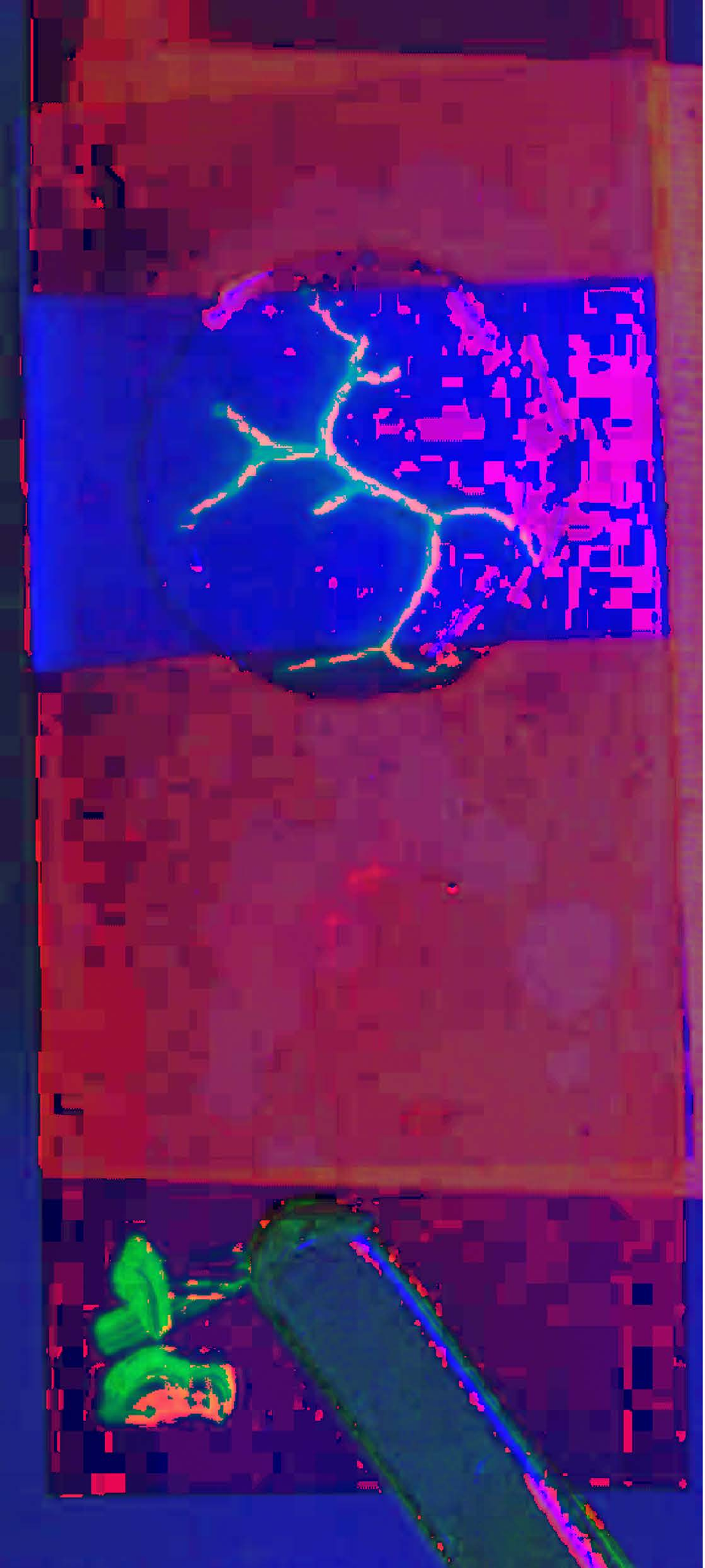}}
    \subfigure[]{\includegraphics[scale=0.45]{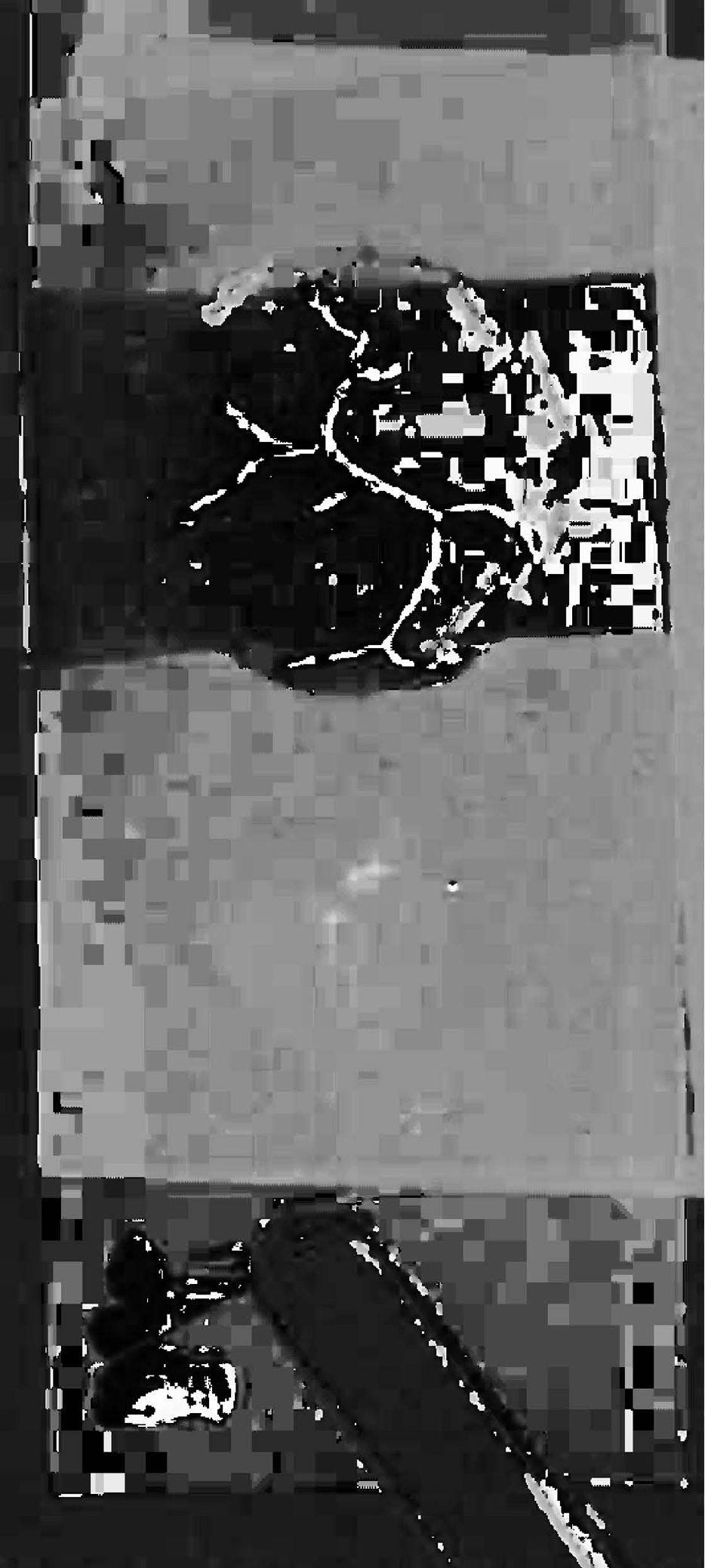}}
    \subfigure[]{\includegraphics[scale=0.45]{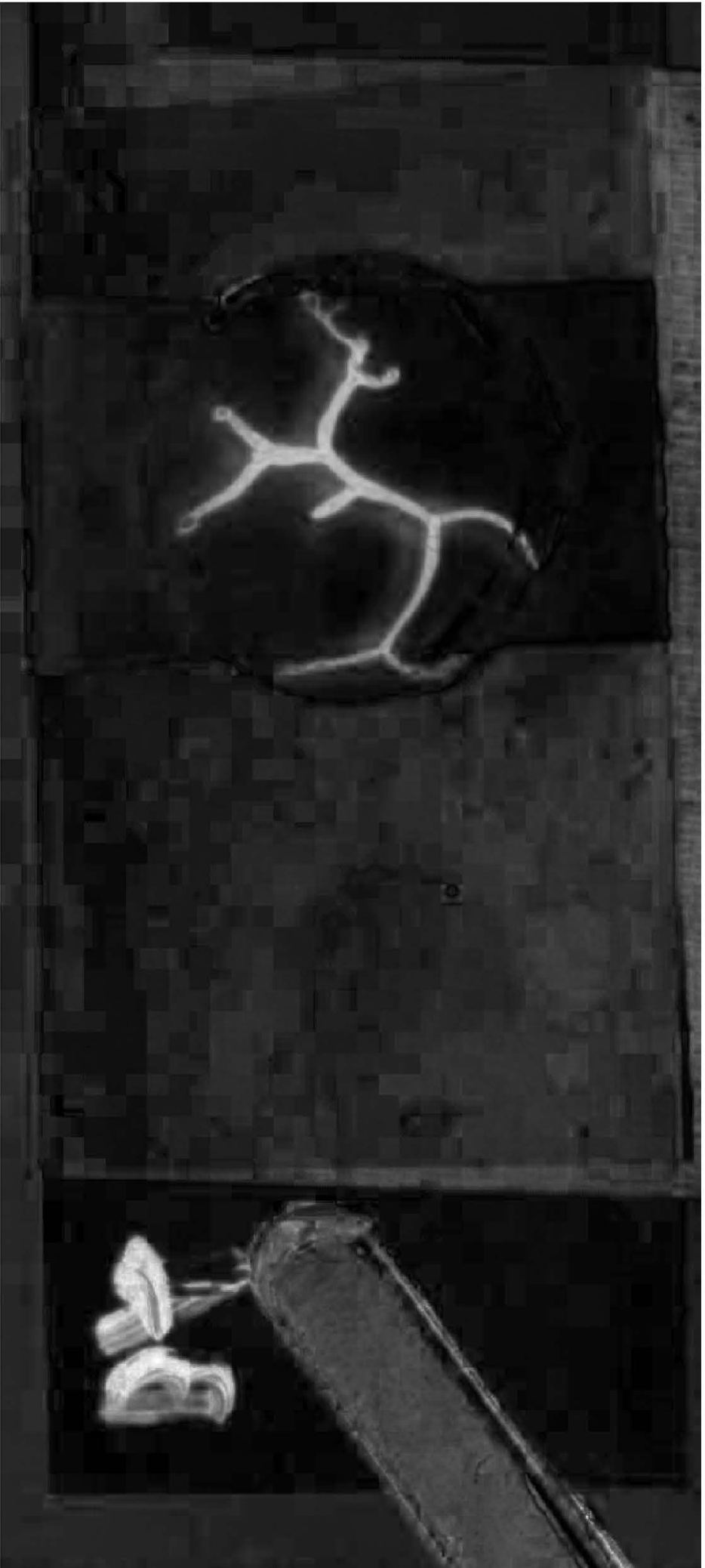}}
    \subfigure[]{\includegraphics[scale=0.45]{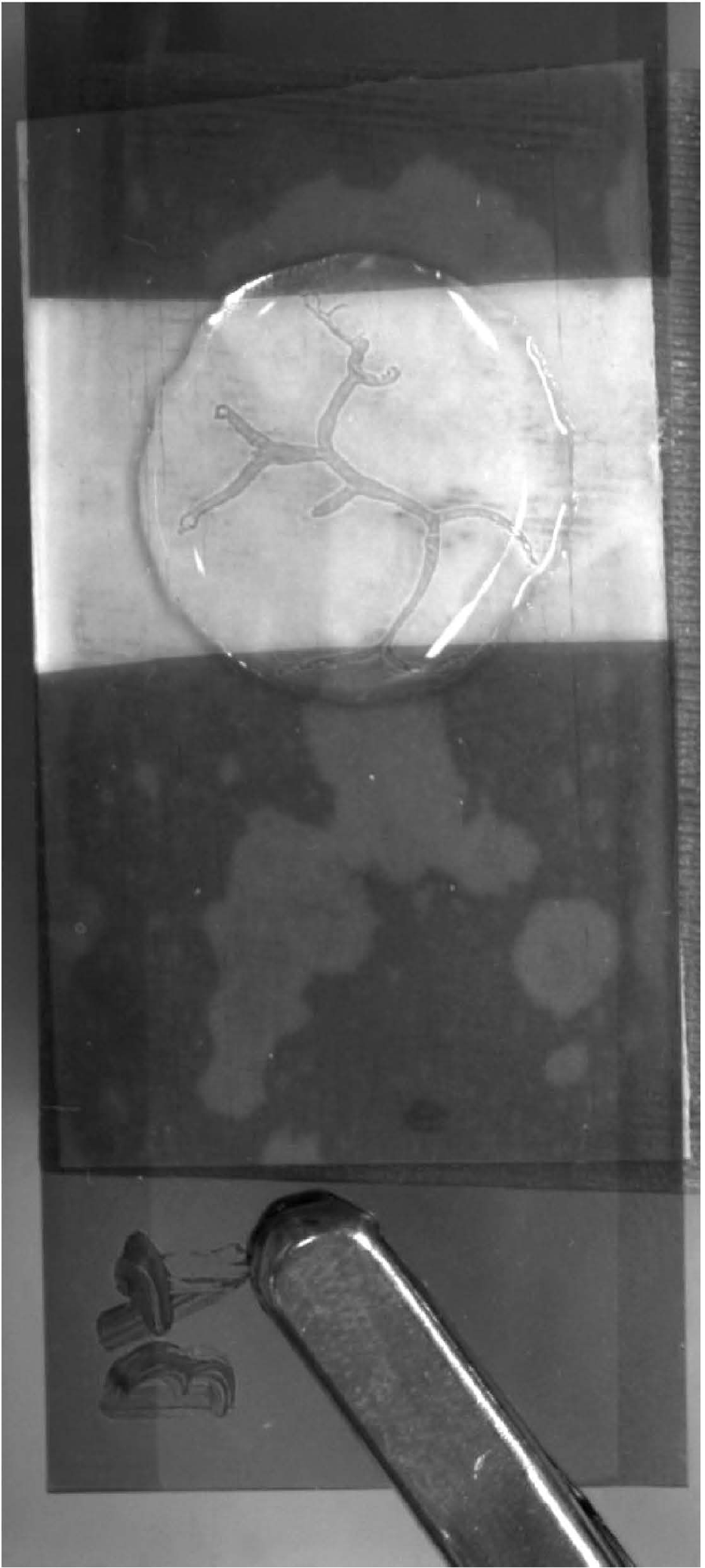}}
    \subfigure[]{\includegraphics[scale=0.47]{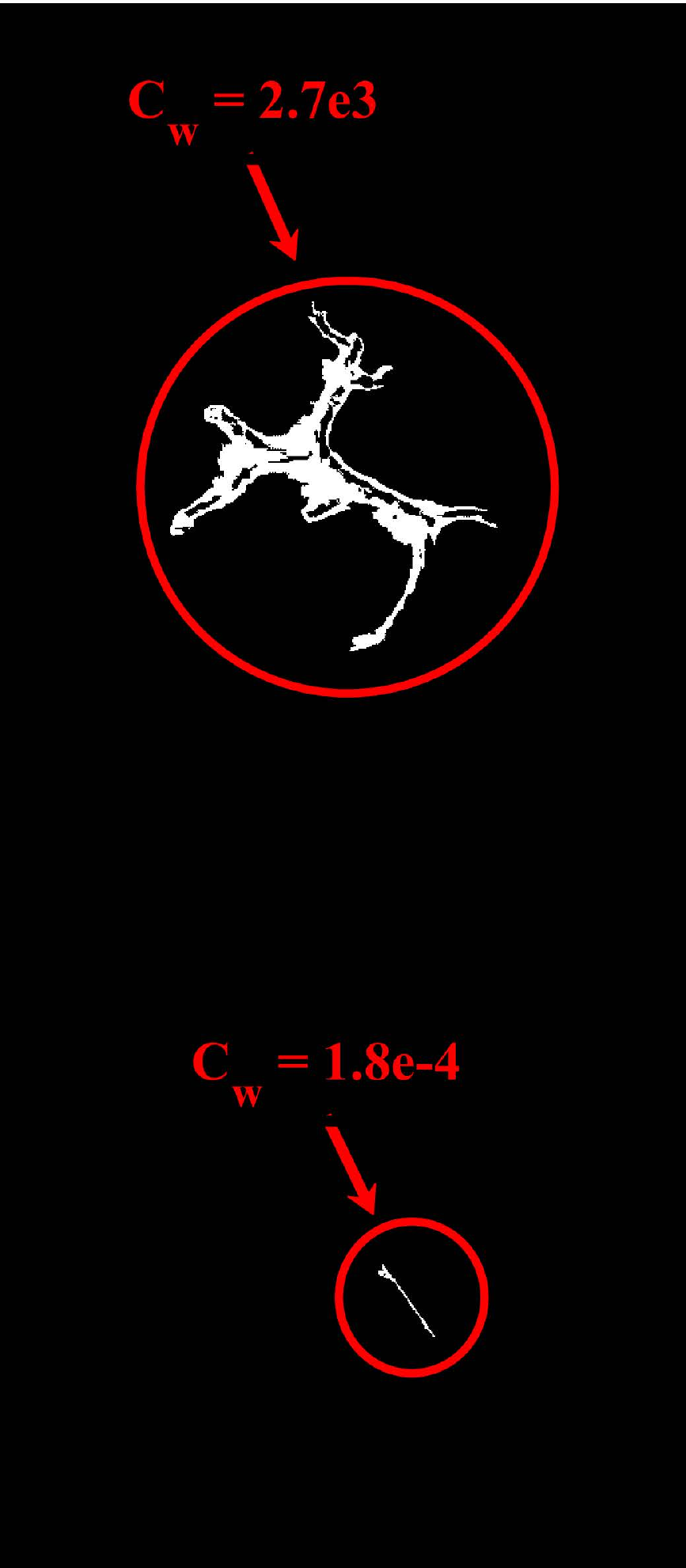}}
    \caption{Basics of droplet image processing.
    (ab)~Image of a droplet in RGB~(a) and HSV~(b) colour spaces. (cde)~Visualisation of the image in Hue~(c), Saturation~(d), and Value~(e) channels, respectively. (f)~Results of applying the segmentation method to Saturation channel in a Binary form. Artefacts are evident in (c), and the Value channel (e) is not suitable for extracting ROI due to the high range of lightness which increases the similarity in different regions. In (f), there are two regions with white pixels surrounding by black pixels where the $C_{w1} = 2.7e3$ and $C_{w2} = 1.8e-04$.}
    \label{fig:M1}
\end{figure}

Droplets have tree-like structures. Therefore, finding the best combination of nodes and edges, which match to the shape, is the key to complexity analysis. In our experiments, the region of interest (ROI) encompasses decanol droplets, and converting the colour space from RGB to HSV could help to highlight ROI. Figure~\ref{fig:M1}a--b shows a droplet in RGB, HSV colorspaces and Fig.~\ref{fig:M1}c--e shows the droplet in Hue, Saturation, and Value channels, respectively. As seen in Fig.~\ref{fig:M1}c, ROI contains artefacts and Fig.~\ref{fig:M1}(e) shows that Value channel, which is defined as the largest component of a colour, cannot bring rich information to extract ROI. Therefore, Saturation channel is taken into account, and the adaptive thresholding method which was proposed in~\cite{dehshibi2017hybrid,ramin2012counting} is applied to perform the segmentation. While some outlines remain to the output image, a post-processing algorithm is then utilised to extract ROI from Fig.~\ref{fig:M1}f.

In the post-processing phase, the white pixels were counted and divided by the size of the image to calculate the normalised coverage area ($C_{w}$). Then the region with the maximum $C_{w}$ was selected as ROI. While the final objective is to construct a graph that matches to the decanol droplet, ROI was enhanced by applying morphological operators to the image. Then, Algorithm~\ref{alg:1} was applied to find the $(x, y)$ coordinates of critical points.

\begin{algorithm}[H]
	\caption{Converting droplet image into graph.} \label{alg:1}

	\hspace*{\algorithmicindent} \textbf{Input:} \\
	\hspace*{3em} $BW: \gets$ Binary image of ROI . \\
	
	\hspace*{\algorithmicindent} \textbf{Output:} \\
	\hspace*{3em} $CP \gets$ coordinates of critical points.\\
	
	\begin{algorithmic}[1]
		\State Removing all minima from $BW$ which are not connected to the image border.
		    \begin{itemize}
				\justifying
				\item[] $SE$ is a disk-shaped structuring element with a radius of 4.
				\item[] $X_{k}=(X_{k-1} \oplus SE) \cap BW^{'}, \quad k = 1,2,3, \cdots $
		    \end{itemize} 
		 \Comment{\textcolor{darkgray}{COMMENTS}} \newline \Comment{\textcolor{darkgray}{$\% X$ is a set of pixels iteratively get updated, where $X_{0} = p$ and $p$ is a single pixel.}} \newline \Comment{\textcolor{darkgray}{$\% BW^{'}$ is the complement of $BW$.}} \newline \Comment{\textcolor{darkgray}{$\% \oplus$ is the dilation operator.}}
		\State Extracting the ROI boundary.	
		    \begin{itemize}
				\justifying
				\item[] $\Delta BW = BW - (BW \ominus SE)$.
		    \end{itemize}
		 \Comment{\textcolor{darkgray}{COMMENTS}} \newline \Comment{\textcolor{darkgray}{$\% \ominus$ is the erosion operator.}}
		 \State Solving the Eikonal equation~\cite{sethian1999level} by computing the distance transform by
		    \begin{itemize}
				\justifying
				\item[] $\nabla (DT)=1$
				\item[] $DT(p)=\min_{q \in \Delta BW} \parallel p-q \parallel_{2}, \quad p \in X_{k}$
		    \end{itemize}
		 \Comment{\textcolor{darkgray}{COMMENTS}} \newline \Comment{\textcolor{darkgray}{$\% \parallel \cdot \parallel_{2}$ is the Euclidean distance.}} \newline \Comment{\textcolor{darkgray}{$\% DT_{0} = 0$ which is the initial $DT$.}}
		 \State Finding two sets of points' coordinates on $\Delta BW$ that satisfy:	
		    \begin{itemize}
				\justifying
				\item[] $S(\Delta BW, U_{i}) = \{(x,y)│ \max(U_{x+1,y}-U_{x,y}, U_{x,y+1}-U_{x,y})> \tau \}, \quad i=1,2$.
		    \end{itemize}
		 \Comment{\textcolor{darkgray}{COMMENTS}} \newline \Comment{\textcolor{darkgray}{$\% \tau$ is a pixel level threshold where boundary details are shorter than it.}} \newline \Comment{\textcolor{darkgray}{$\% U$ is a set of coordinates in which the $\parallel \alpha - q \parallel_{2}$ for all $q \in \Delta BW$ are equal. Here, $\alpha$ is a fixed starting point.}} \newline \Comment{\textcolor{darkgray}{$\%$ The starting points $(\alpha)$ of $U_{1}$ and $U_{2}$ are not the same.}}
		 \State The skeleton is given by:	
		    \begin{itemize}
				\justifying
				\item[] $\delta BW = \min \{S(\Delta BW, U_{1}), S(\Delta BW, U_{2})\}$.
		    \end{itemize}
		 \State Calculating the Reeb graph (RG)~\cite{biasotti2000extended} on $\delta BW$ set, and mapping it to $\Delta BW$ using~\cite{lazarus1999level}.	
		    \begin{itemize}
				\justifying
				\item[] $CP=\{(x,y) \in RG | \ulcorner \delta BW_{(x,y)}=0 \quad \mathrm{and} \quad e_{(x,y)}=\mathbf{TRUE}\}$
		    \end{itemize}
		 \Comment{\textcolor{darkgray}{COMMENTS}} \newline \Comment{\textcolor{darkgray}{$\% \ulcorner$ is the gradient.}} \newline \Comment{\textcolor{darkgray}{$\% e_{(x,y)}=\mathbf{TRUE}$ reveals that edges of the graph represent connections.}}
		 \Return{$CP$}
	\end{algorithmic}
\end{algorithm}

\begin{figure}[!tbp]
    \centering
    \subfigure[]{\includegraphics[scale=0.805]{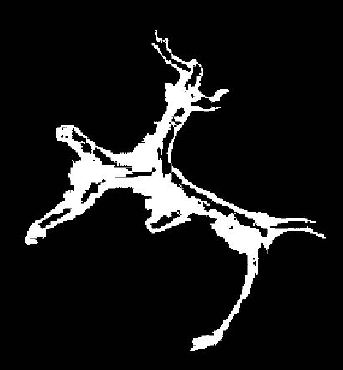}}
    \subfigure[]{\includegraphics[scale=0.75]{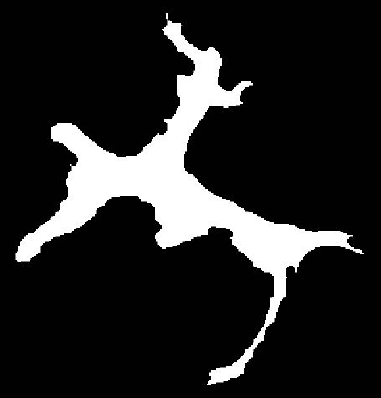}}
    \subfigure[]{\includegraphics[scale=0.395]{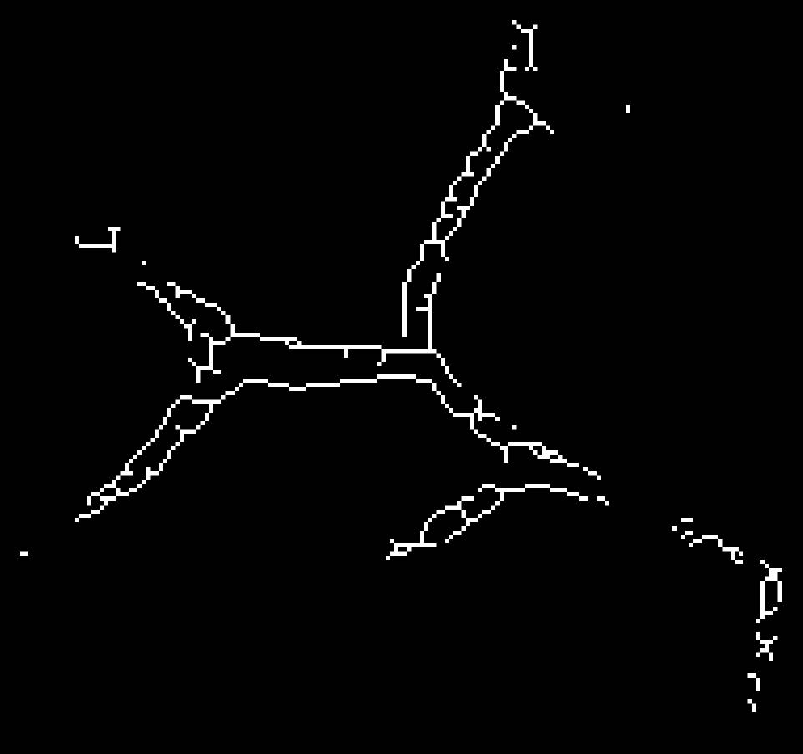}}
    \subfigure[]{\includegraphics[scale=0.4]{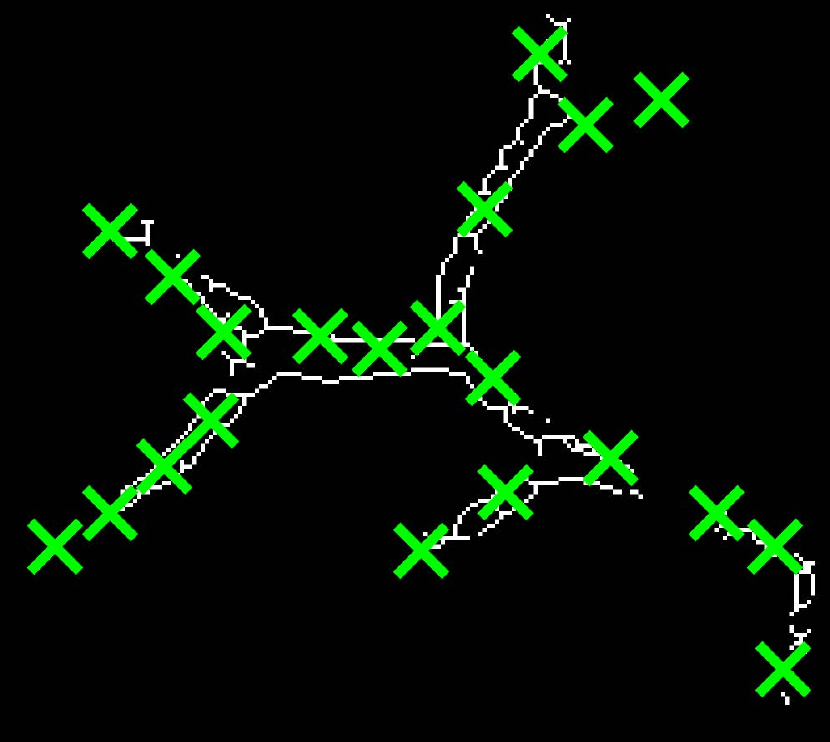}}
    \caption{Converting a droplet to a graph. 
    (a)~ROI, $BW$. 
    (b)~Results of applying Step 1, $X_8$. 
    (c)~Results of applying Steps 2-5. 
    (d)~Visualization of critical points' 
    $(x, y)$--coordinates on the skeleton of Decanol.}
    \label{fig:M2}
\end{figure}

Results of applying Algorithm~\ref{alg:1} to ROI are shown in Fig.~\ref{fig:M2}.

In terms of complexity analysis, four measurements were calculated based on the image and the extracted critical points from ROI which are: (1)~normalized coverage area, (2)~compressibility, (3)~morphological richness~\cite{adamatzky2000choosey,adamatzky2010generative,taghipour2016complexity}, and (4) entropy of Delaunay triangulation. Normalised coverage area is calculated by counting the number of with pixels in ROI divided by the size of Image ($C_w$). Compressibility expresses as the size of the PNG image in bytes. Morphological richness (MR) is calculated in a local neighbourhood of each pixel with a radius of 3. It is the number of different configurations of each $3 \times 3$ blocks divided by the number of all possible configurations ($2^{9} = 512$). When the MR of all ROIs were calculated, we focused on two sub-measurements to provide a better inference. The power spectrum~\cite{gholami2018complexity} of the entropy of MRs and distances between each pair of MRs were calculated to make the complexity analysis sensible. Finally, a Delaunay triangulation~\cite{adamatzky2005reaction} was created using the CP set, and we calculated the entropy of the adjacency matrix.

\section {Results}

\begin{figure}[!tbp]
    \centering
   \subfigure[]{\includegraphics[width=0.49\linewidth]{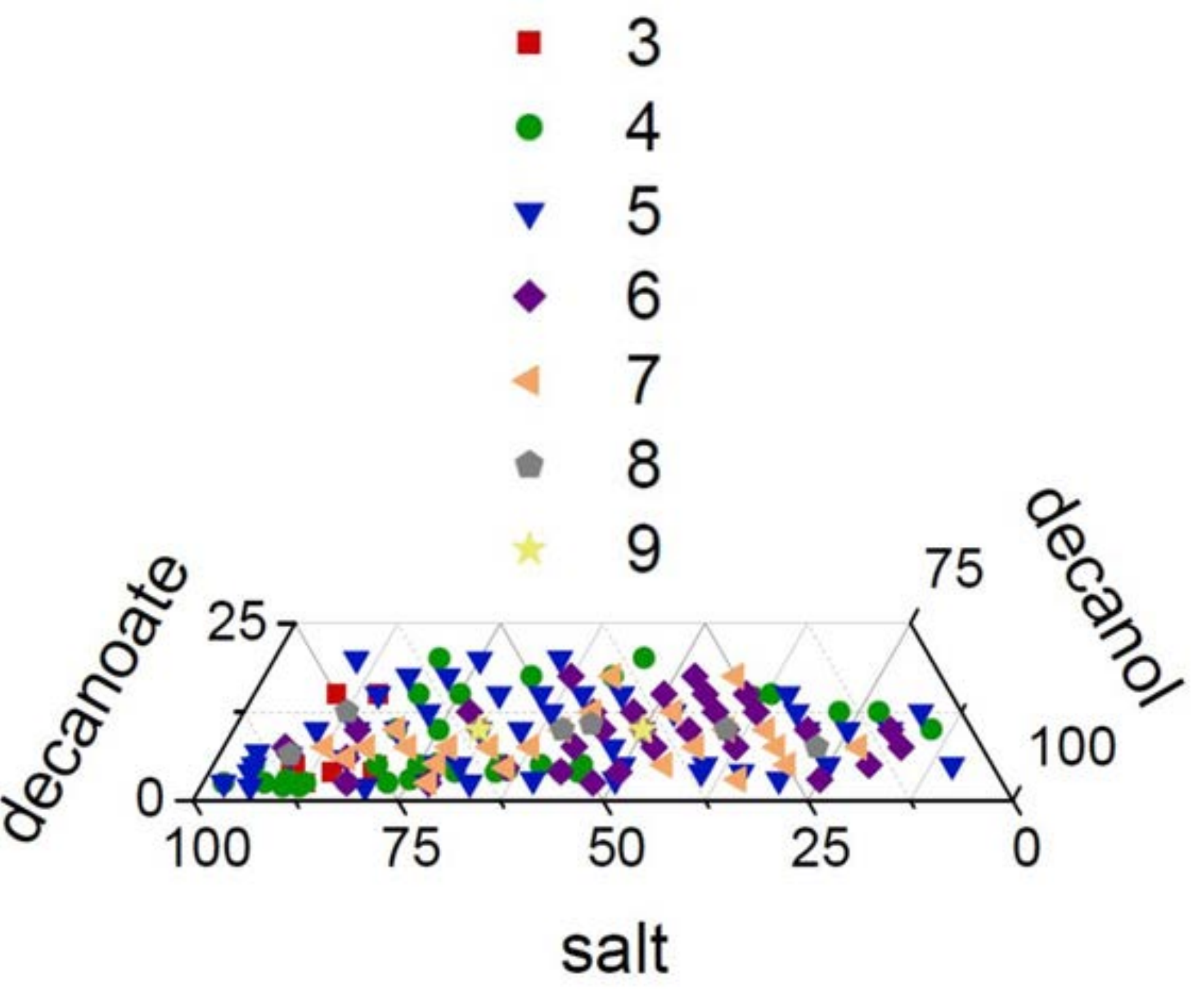}}
      \subfigure[]{\includegraphics[width=0.49\linewidth]{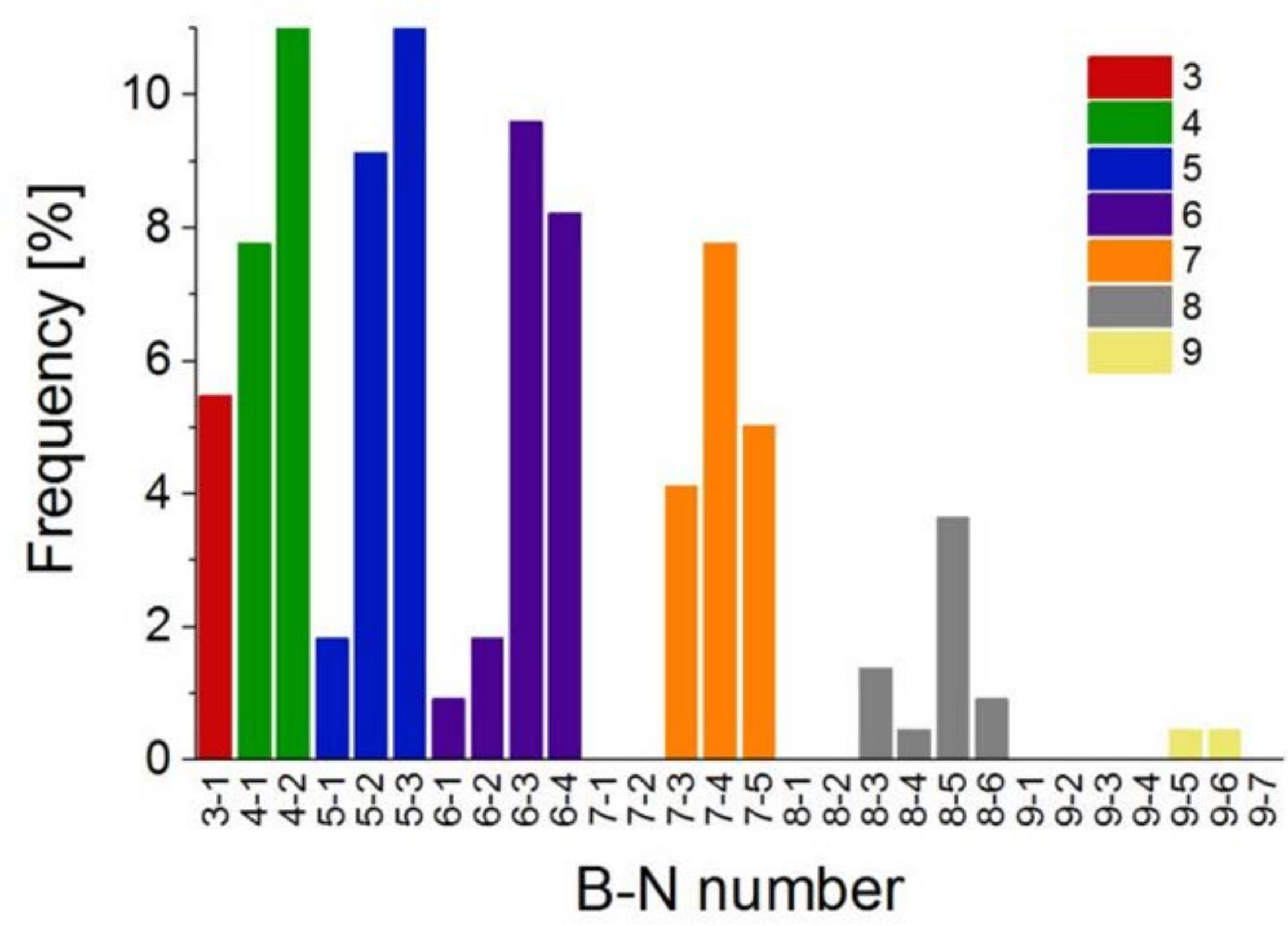}}
    \caption{(a)~The number of branches of tentacular structures for various molar ratios of decanol–-decanoate–--salt. 
    (b)~Percentage of droplets in accordance of the B--N number, i.e. the ratio of the free ends of tentacular structures to the number of nodes inside the branching structure. Evaluated from 220 images.}
    \label{fig:1}
\end{figure}

We have performed several  experiments by changing the molar ratio of decanol--decanoate--salt, and evaluated the number of branches and nodes of the largest tentacular structure before its disintegration into more parts (see Fig.~\ref{fig:1}). It has been found that over 25\% of droplets developed has five branches (i.e. five free ends of the structure).  However, we found that the chemical composition of the system does not affect structure's nodes and branches. It is worth mentioning that this evaluation is related to a comparison of 220 maximal branching structures before their disintegration. Therefore, it does not reflect the evolution of the structures in time and does not provide any indications on future evolution of the droplet after the disintegration of the base structure.

\begin{figure}[!tbp]
    \centering
\includegraphics[width=0.8\linewidth]{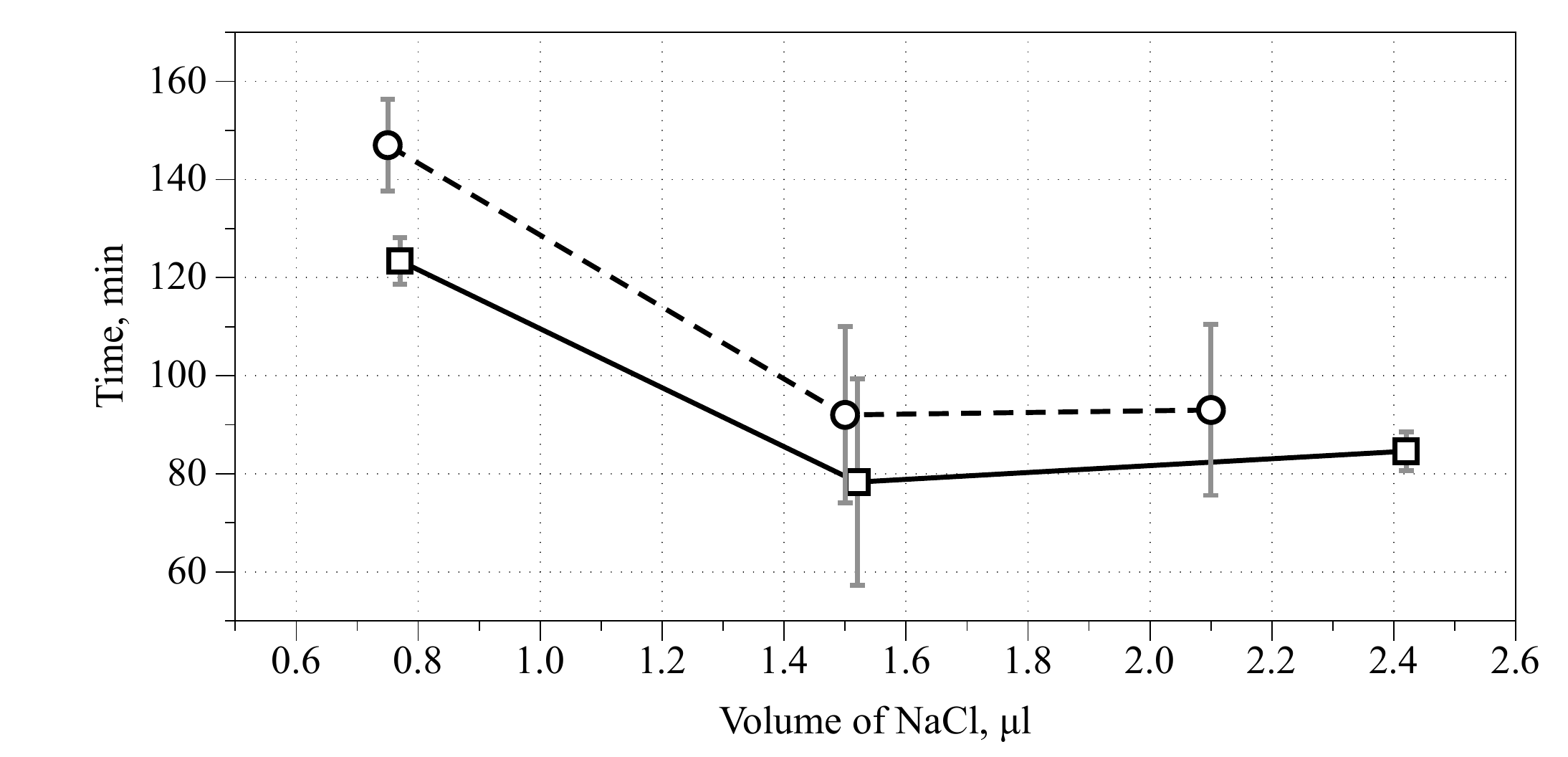}
    \caption{The dependence of the start of pattern formation on an applied voltage (0~DVC  shown by $\circ$, 2~DVC by $\Box$) and an amount of added sodium chloride.}
    \label{fig:C3}
\end{figure}

Further we focused on comparison of experiments without and with electric field. Different amounts of sodium chloride solution (0.75~$\mu$l, 1.5~$\mu$l and 2.4~$\mu$l) were added to 250~$\mu$l of decanoate solution with 2.4~$\mu$l decanol droplet. The time of pattern formation beginning was evaluated in both scenarios - when the applied voltage was off and on. The change was observed between different amounts of added sodium chloride. The decrease of time needed for start of patterning was observed between control droplets and droplets subjected to the electric field (Fig.~\ref{fig:C3}). 

\begin{table}
\centering
\caption{Integral characteristics of droplets: number $d$ of disconnected components in a droplet-graph, number $n$ of nodes in a droplet-graph, LZ complexity $z$ based on JPEG compression size in bytes and normalized coverage $\zeta$.}
\begin{tabular}{l||l|l|l|l}
Parameter &  Square  & Square with DC & Round  & Round with DC  \\\hline \hline
Average $d$	& 1.79 &	1.79	& 1.37 &	1.37 \\
St. dev. $d$ &	0.82 &	0.89 &	0.52 &	0.60 \\
Median $d$ &	2	& 2 &	1 &	1 \\
Max $d$ &	3	& 4 &	2	 & 3 \\ \hline
Average $n$	&	9.59	&	10.97	&	10.25	&	9.5	\\
St dev $n$	&	2.2	&	3.55	&	3.33	&	3.93	\\
Median $n$	&	10	&	11	&	9.5	&	10	\\
Max $n$	&	13	&	19	&	16	&	16	\\ \hline
Average $z$  & 63997 & 74431 & 60288 & 61003 \\  \hline
Average $\zeta$	&	0.71	&	0.71	&	0.51	&	0.47	\\
St dev $\zeta$	& 0.13	    &	0.15	& 0.08		& 0.12		\\
Median $\zeta$	& 0.71		&	0.73	&	0.51	&	0.44	\\
\end{tabular}
\label{table:statistics}
\end{table}

Statistical properties of branching droplets are summarised in Tab.~\ref{table:statistics}. Let us discuss each of the features mentioned in the table in more details.

\begin{figure}[!tbp]
    \centering
\subfigure[]{\includegraphics[scale=0.35]{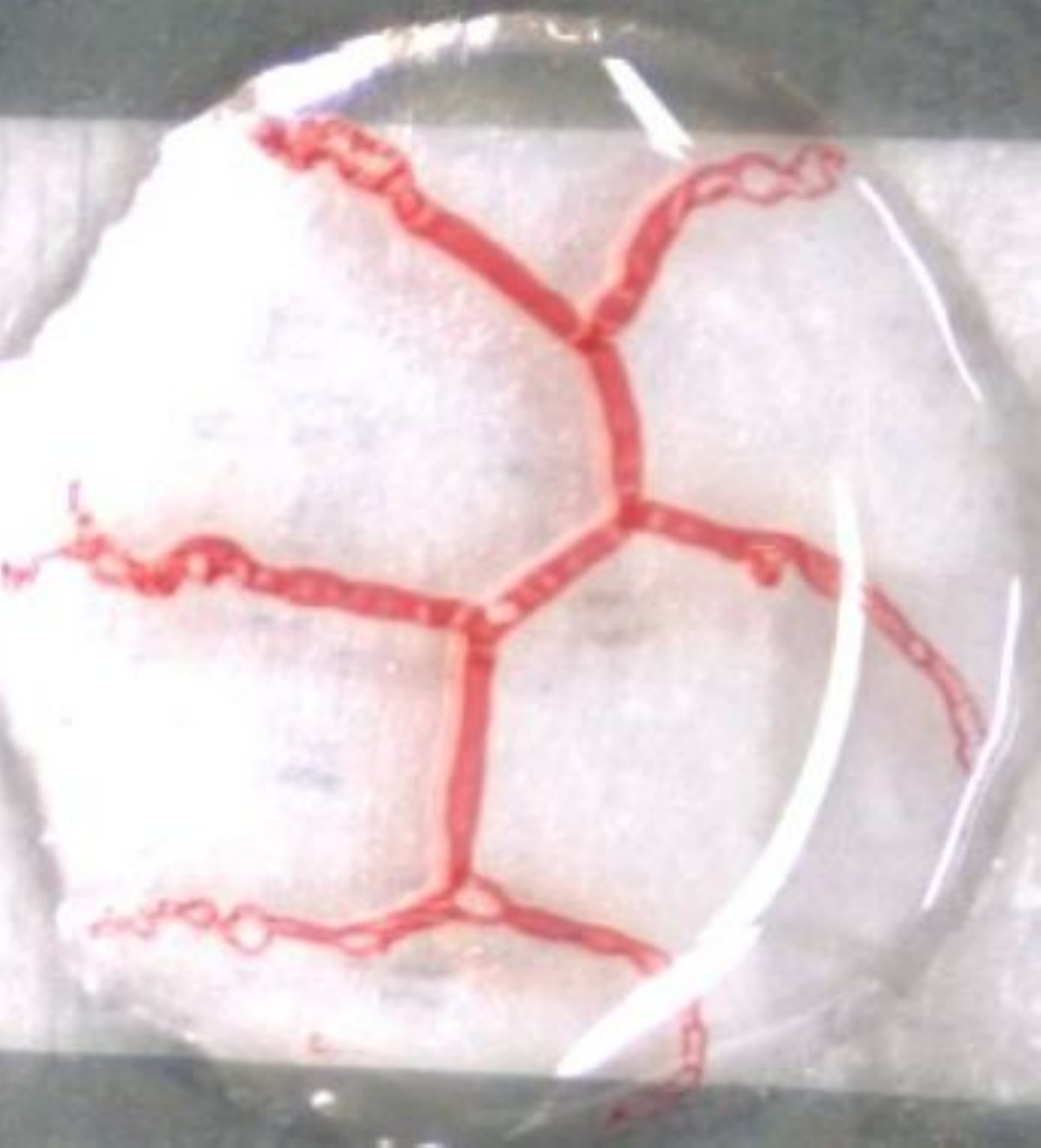}}
\subfigure[]{\includegraphics[scale=0.35]{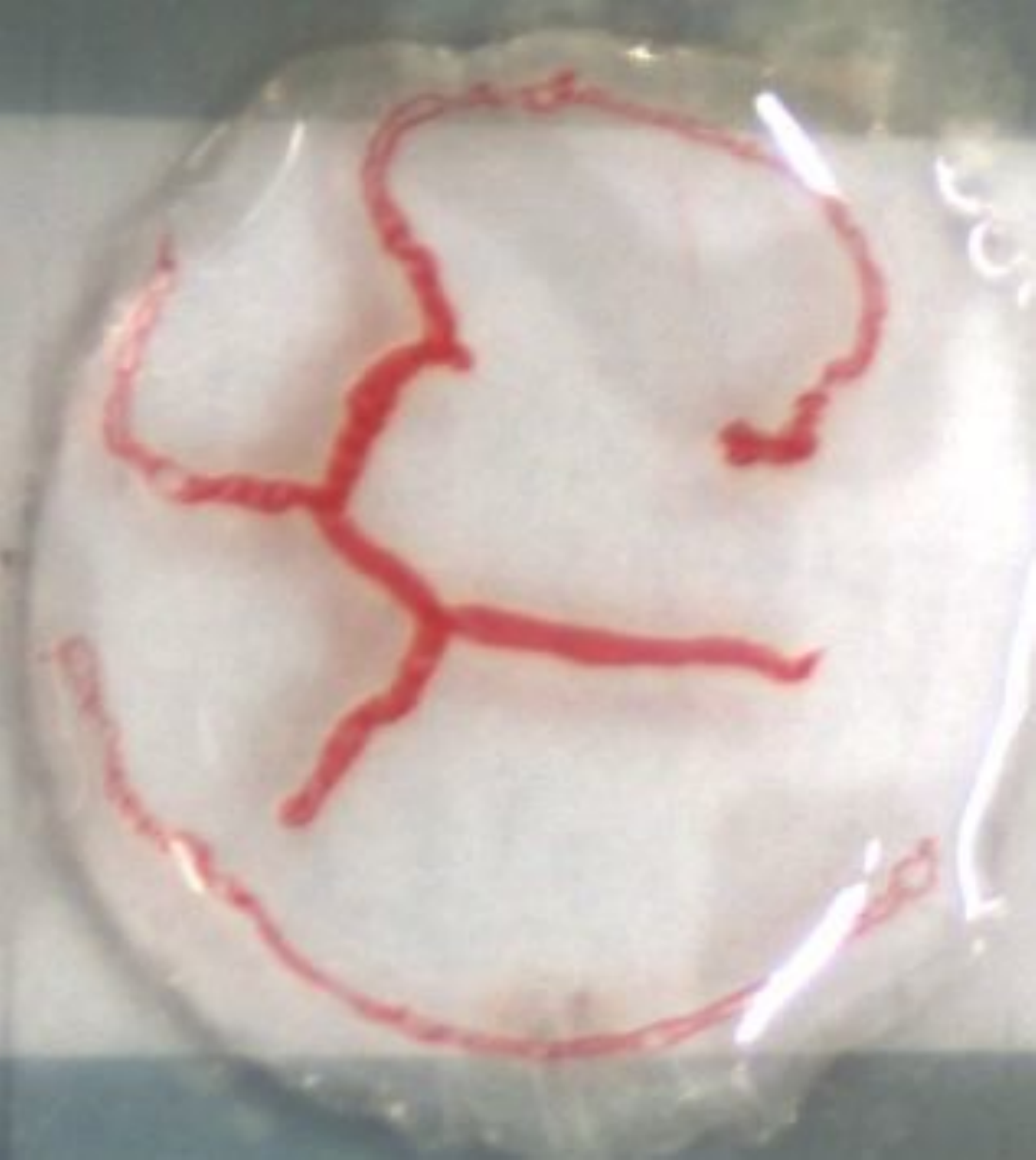}}
\subfigure[]{\includegraphics[scale=0.35]{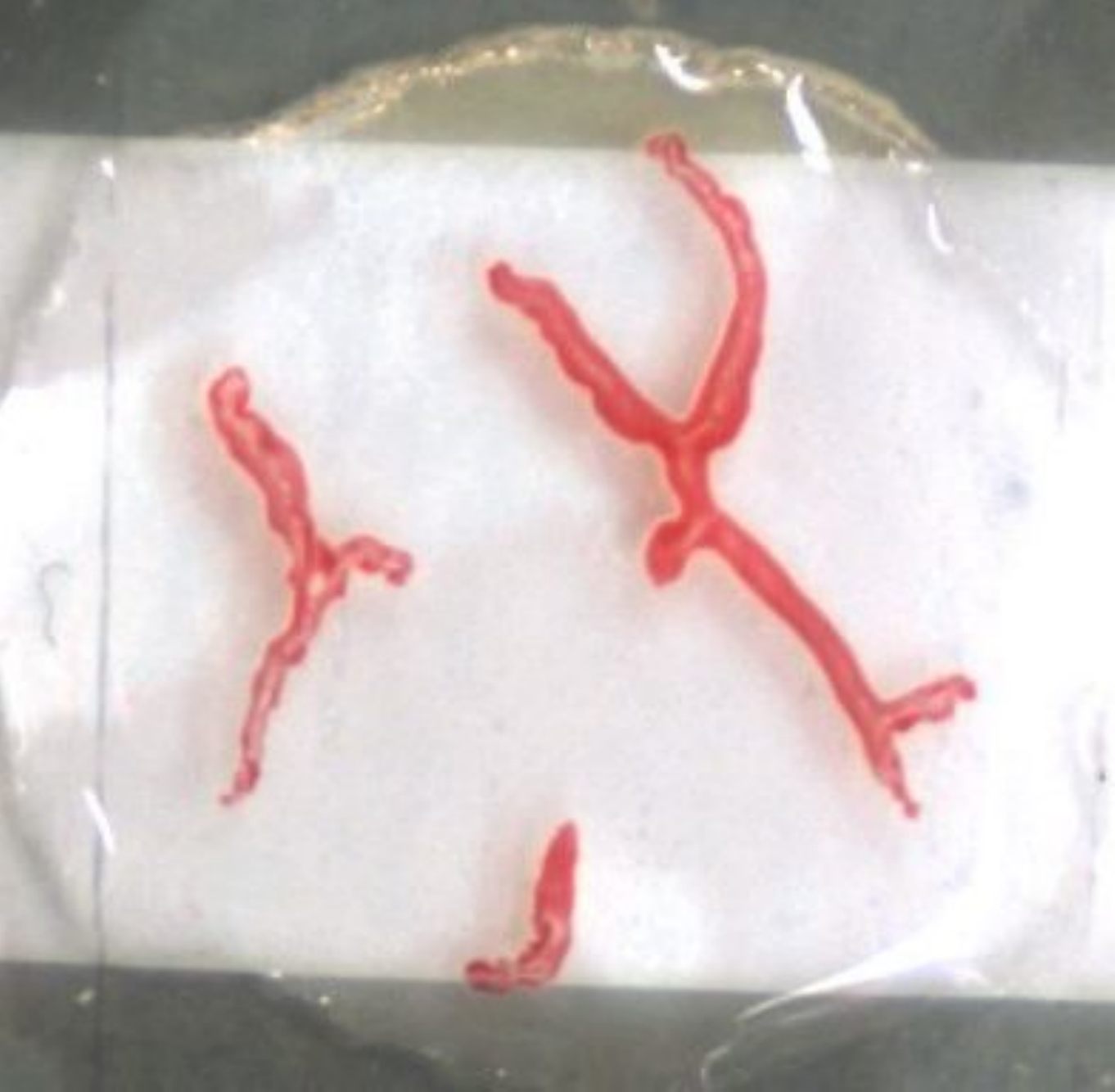}}
\subfigure[]{\includegraphics[scale=0.34]{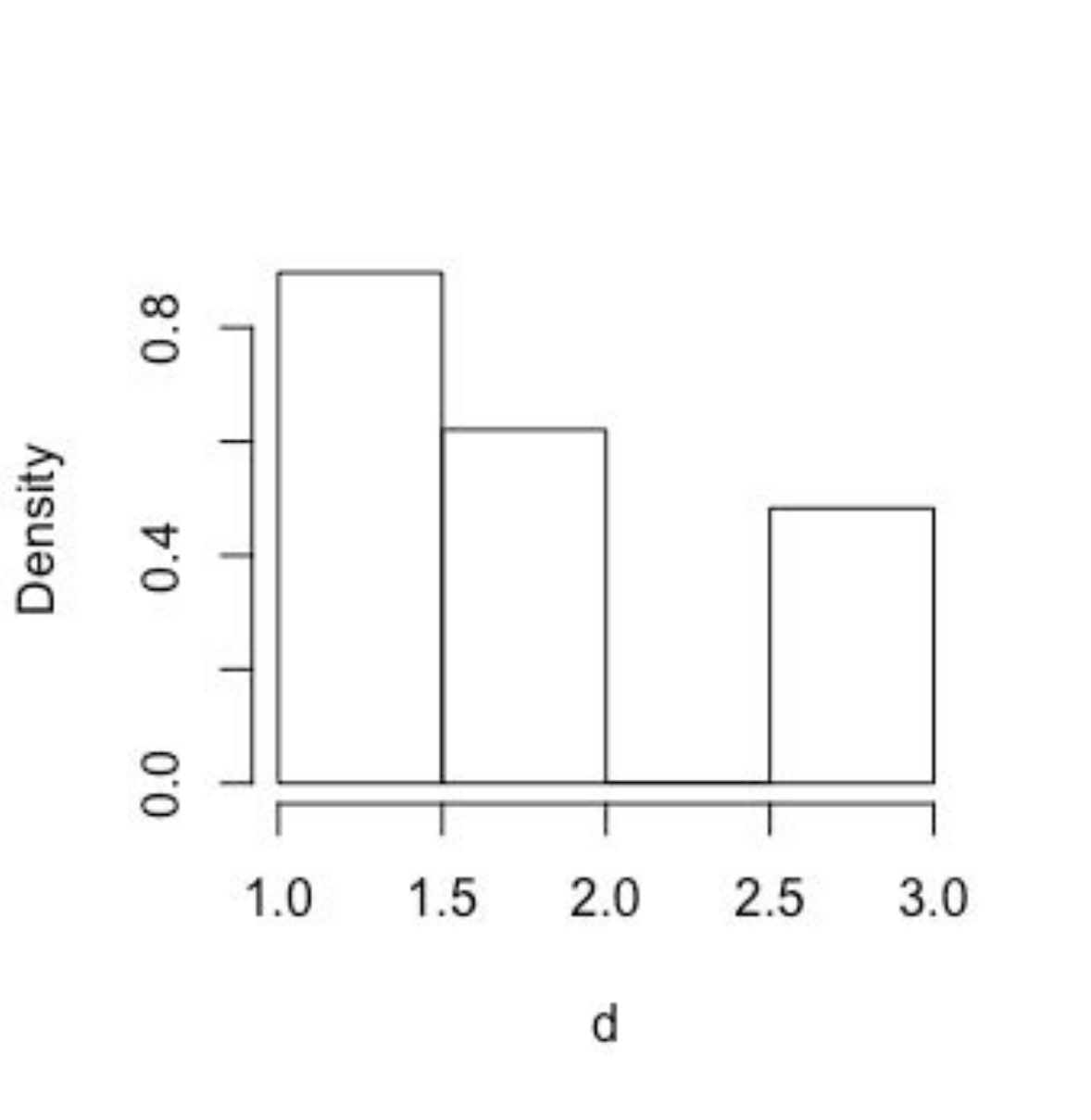}}
\subfigure[]{\includegraphics[scale=0.34]{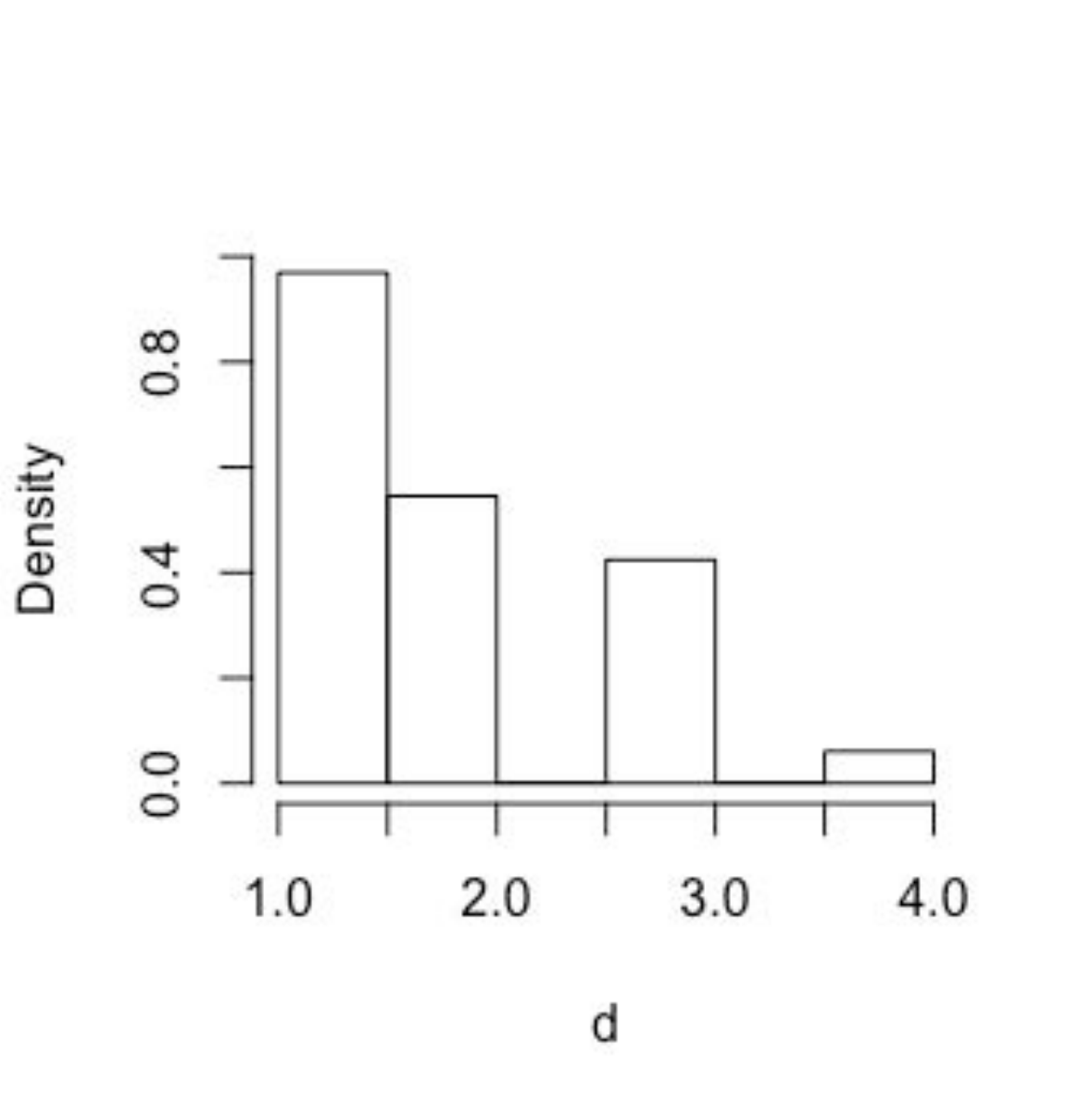}}
\subfigure[]{\includegraphics[scale=0.34]{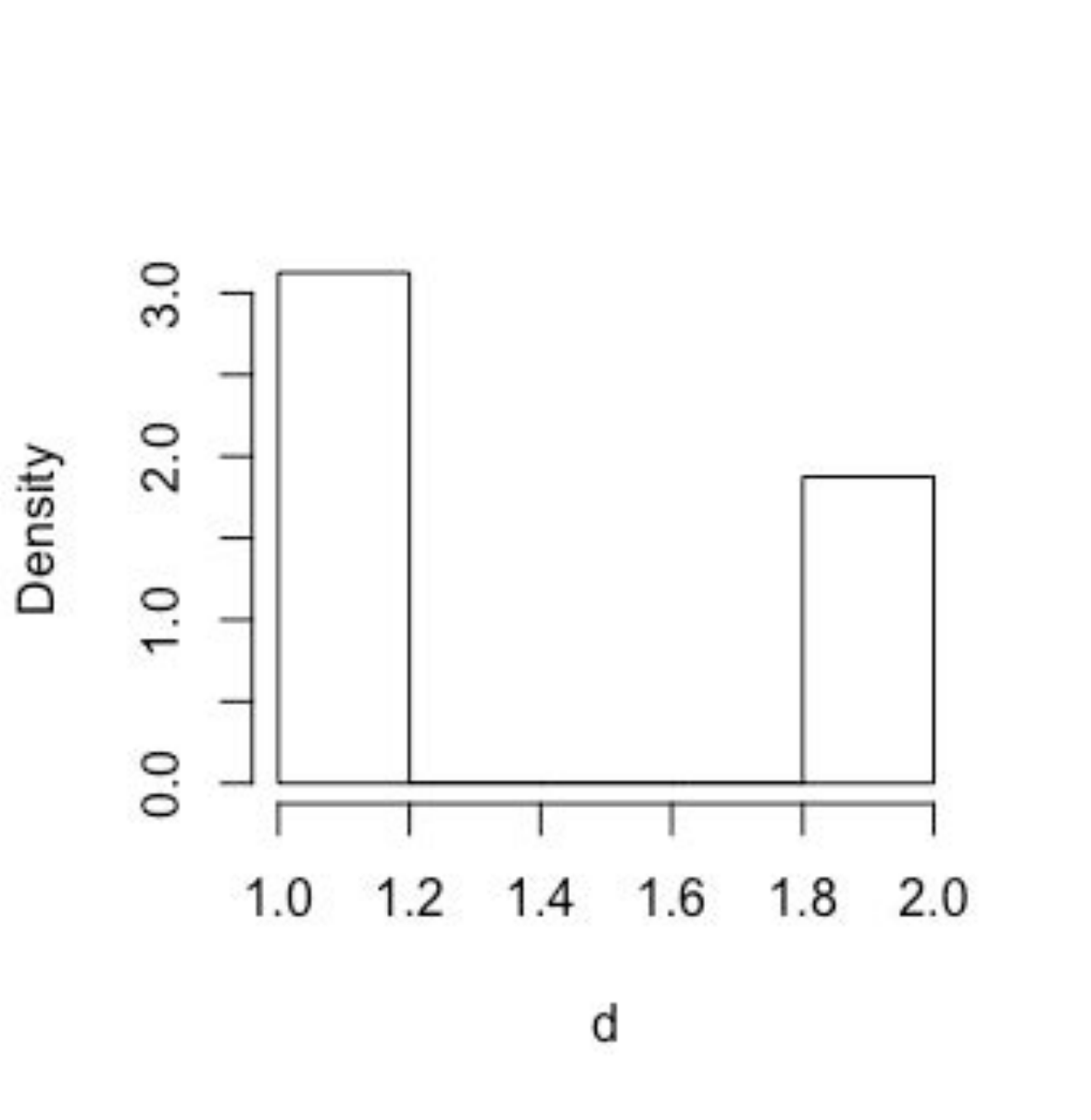}}
\subfigure[]{\includegraphics[scale=0.34]{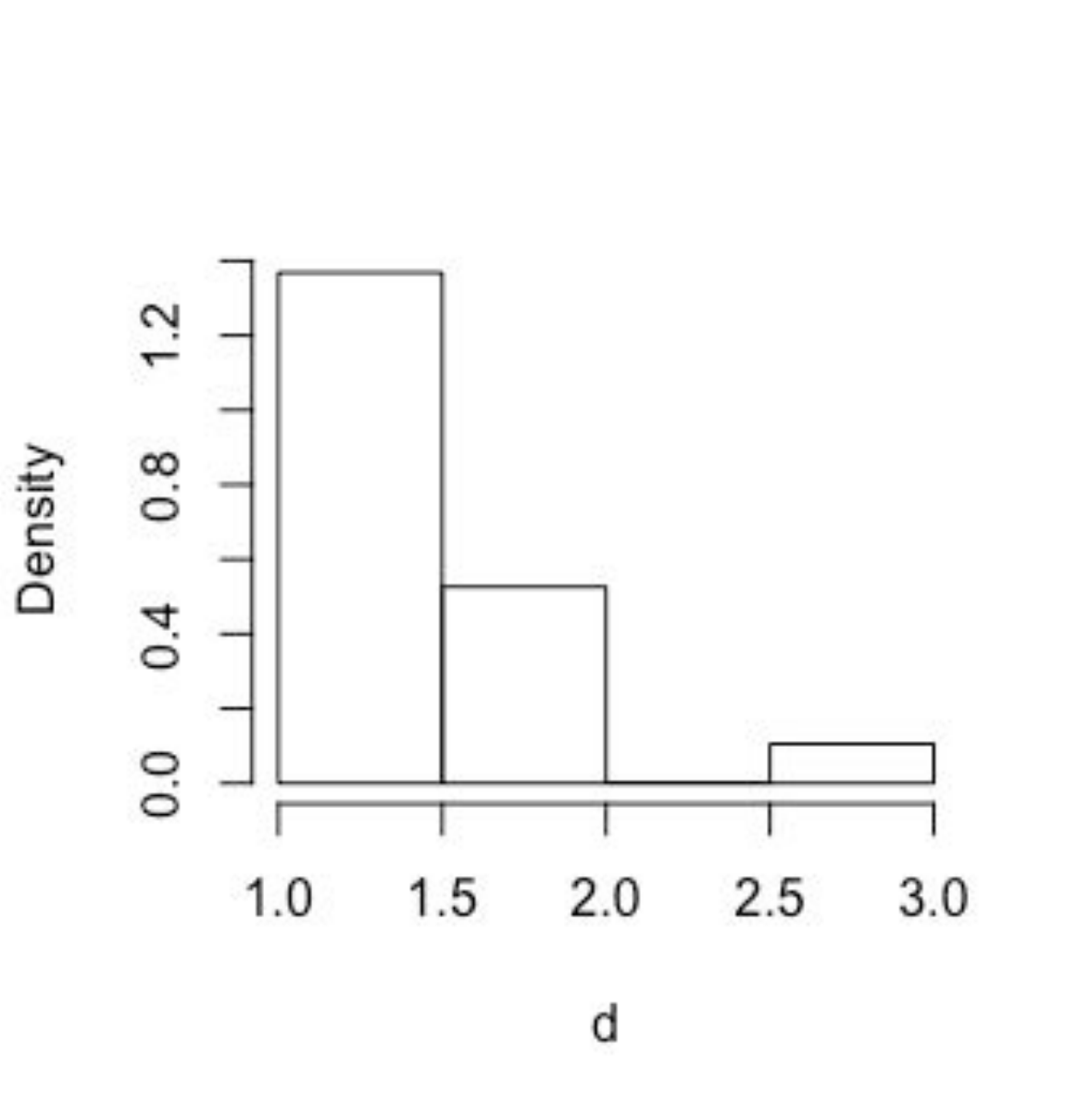}}
    \caption{Examples of branching structures with different number of components~(abc) and density distributions of of numbers of disconnected components (defg). 
    (a)~One connected component. Droplet was subjected to continuous DC of 1.5~V. 
    (b)~Two components. Droplet was subjected to continuous DC of 2~V. 
    (c)~Three components.Droplet was subjected to continuous DC of 1.5~V.
    (d)~Square setup without DC, (e)~Square setup with DC, (f)~Round setup without DC, (g)~Round setup with DC. 
    }
    \label{fig:types}
\end{figure}

\begin{finding}
DC increases a maximal number of disconnected components from 3 in control to 4 in square setup and from 2 to 3 in round setup.
\end{finding}

Graphs representing branching droplets can be connected or disconnected. Disconnected graphs can be several components, see examples in Fig.~\ref{fig:types}abc. Density distributions of number of components are shown in Fig.~\ref{fig:types}d--g. Distributions for droplets without DC shows gaps in the middle, between low and higher numbers of components. Application of DC leads to increase of variability, gaps are filled, and increase of a number of disconnected components.

\begin{figure}[!tbp]
\centering
\subfigure[]{\includegraphics[scale=0.34]{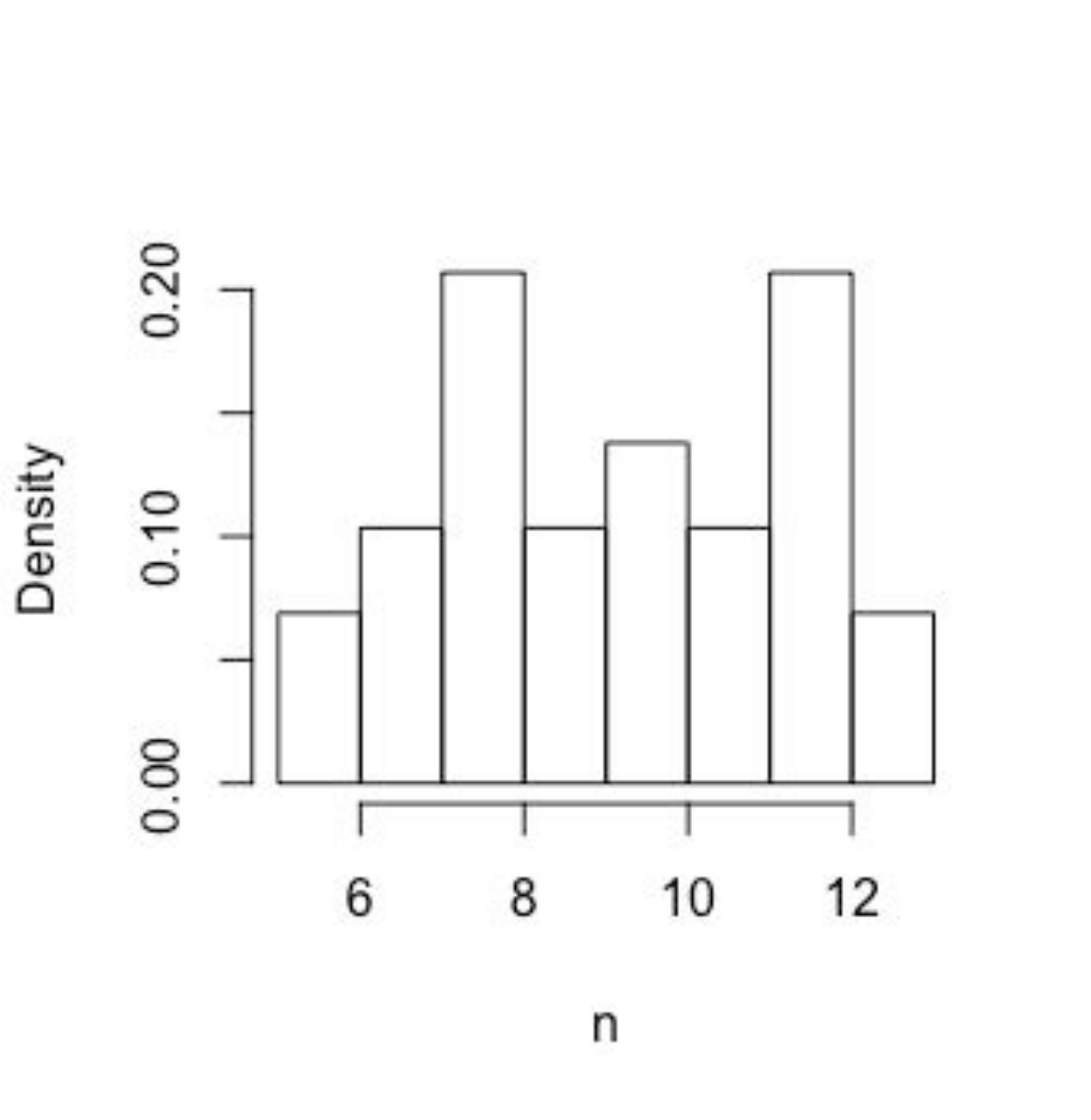}}
\subfigure[]{\includegraphics[scale=0.34]{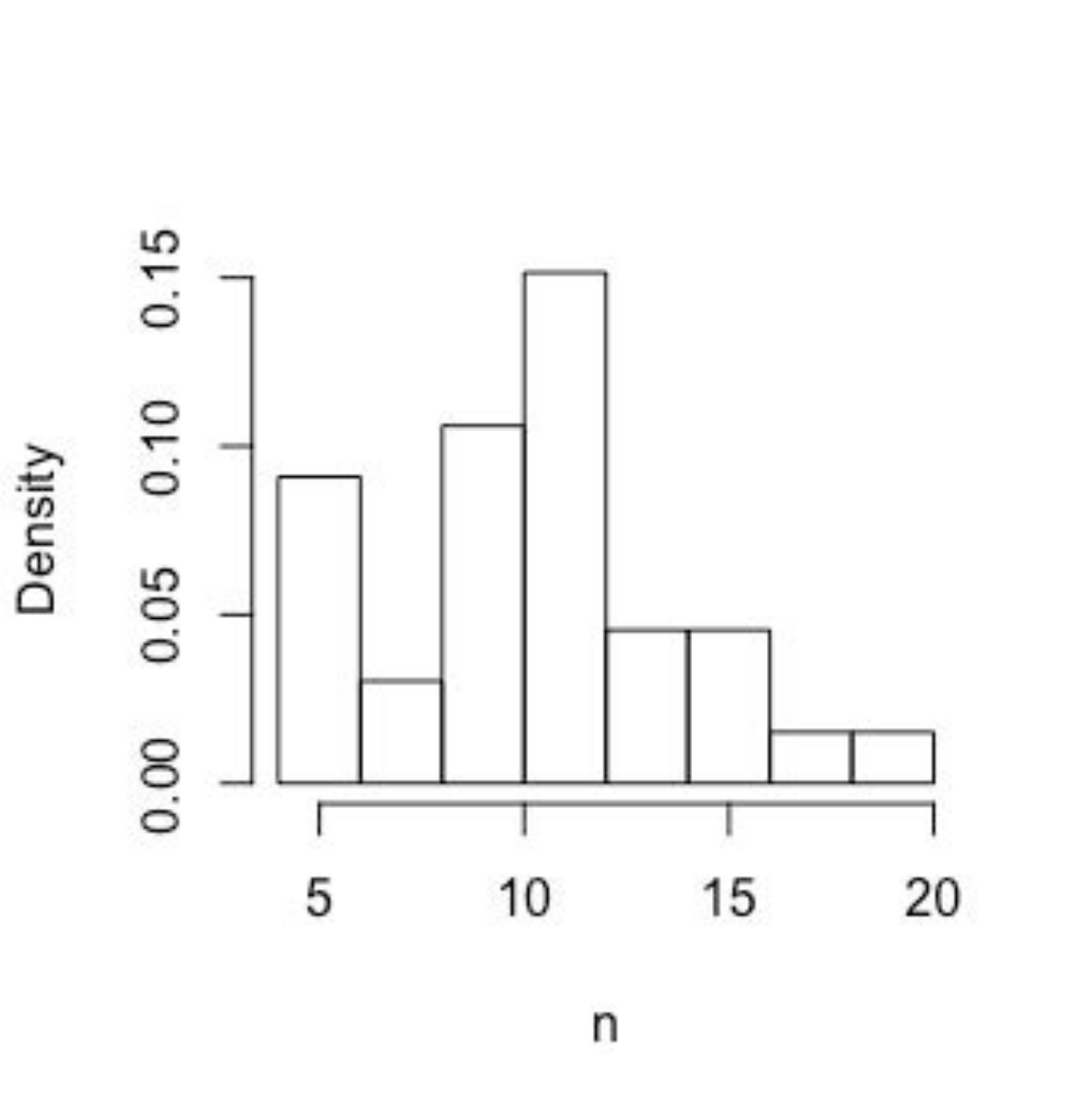}}
\subfigure[]{\includegraphics[scale=0.34]{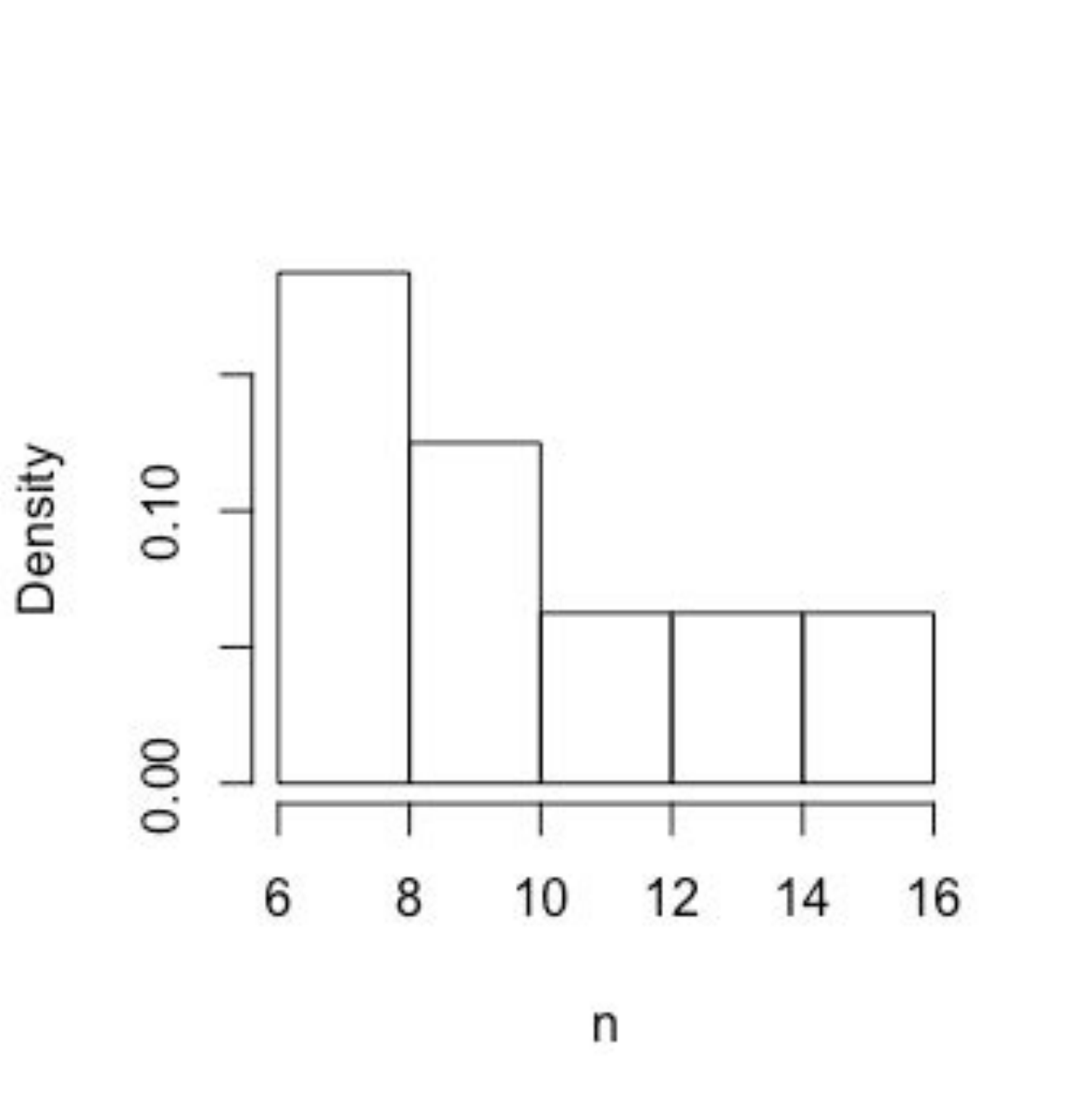}}
\subfigure[]{\includegraphics[scale=0.34]{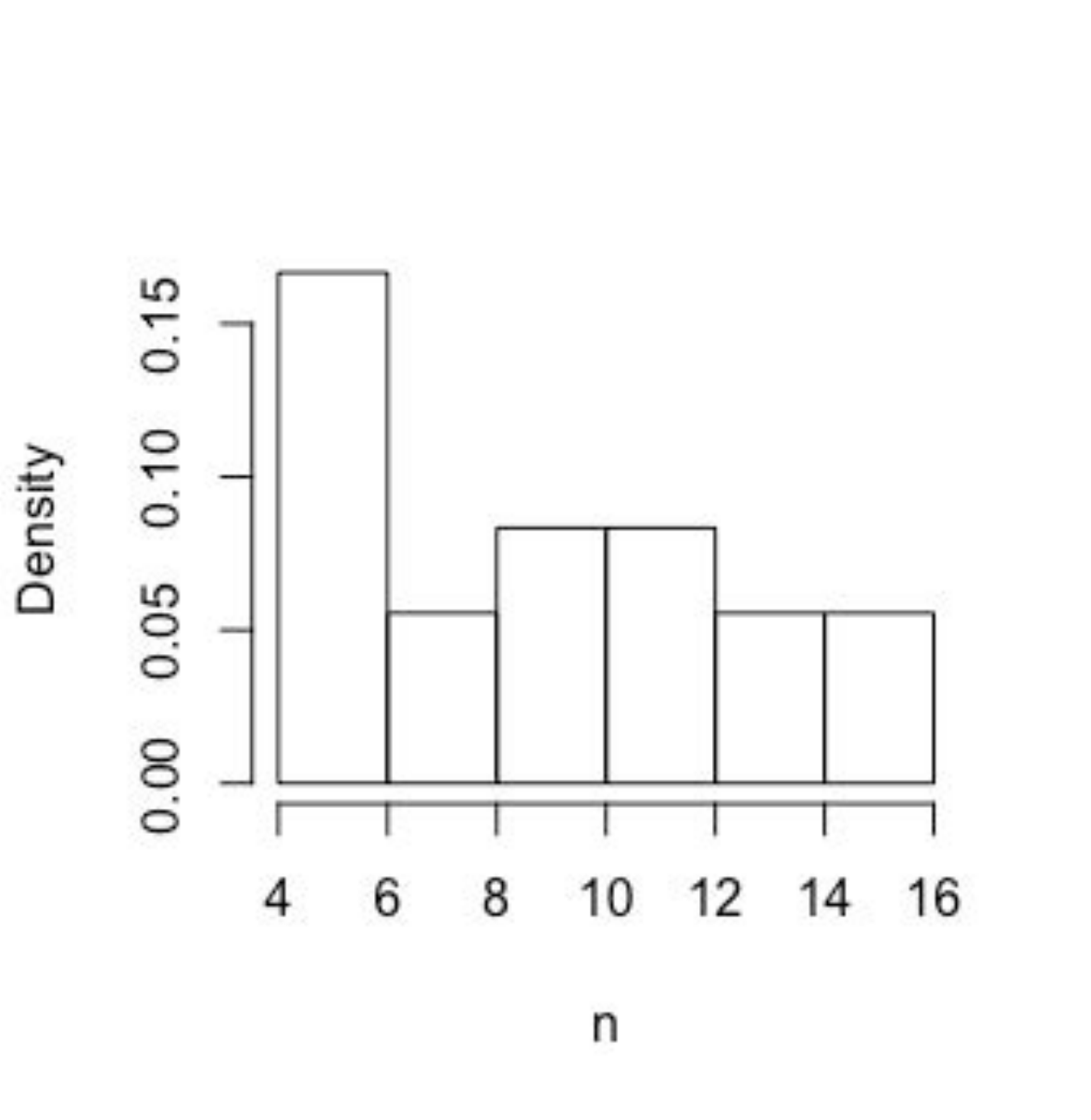}}
\caption{Density distribution of numbers of nodes.  
(a)~Square setup without DC. 
(b)~Square setup with DC. 
(f)~Round setup without DC. 
(g)~Round setup with DC. 
}
\label{fig:distributionNodes}
\end{figure}

\begin{finding}
DC  increases a maximum number of nodes from 13 to 19 (for square); DC also leads to increase of the medium number of nodes.
\end{finding}

This follows from density distributions of a number of nodes in Fig.~\ref{fig:distributionNodes}.

\begin{figure}[!tbp]
\centering
\subfigure[]{\includegraphics[scale=0.3]{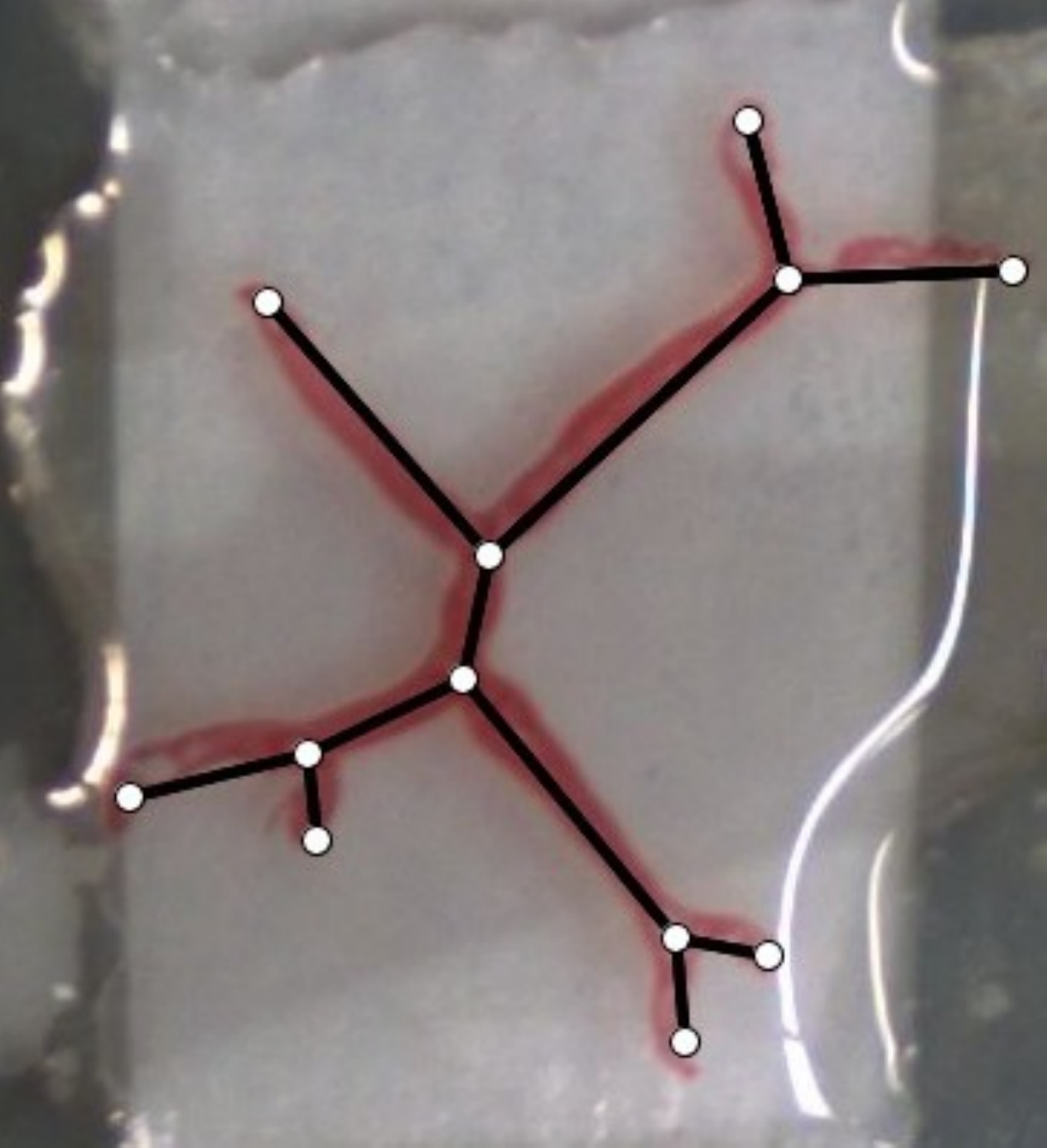}}
\subfigure[]{\includegraphics[scale=0.3]{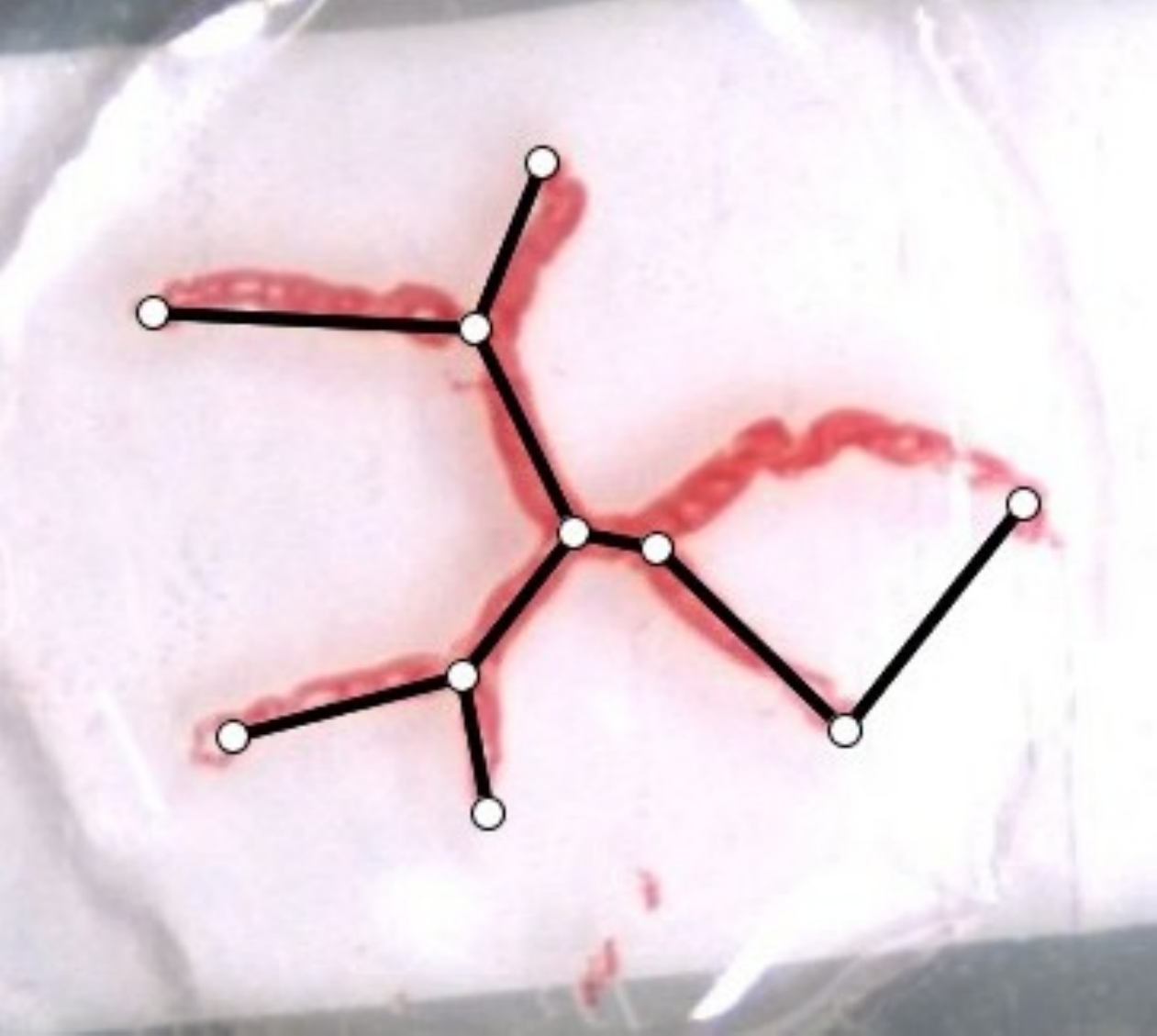}}
\subfigure[]{\includegraphics[scale=0.3]{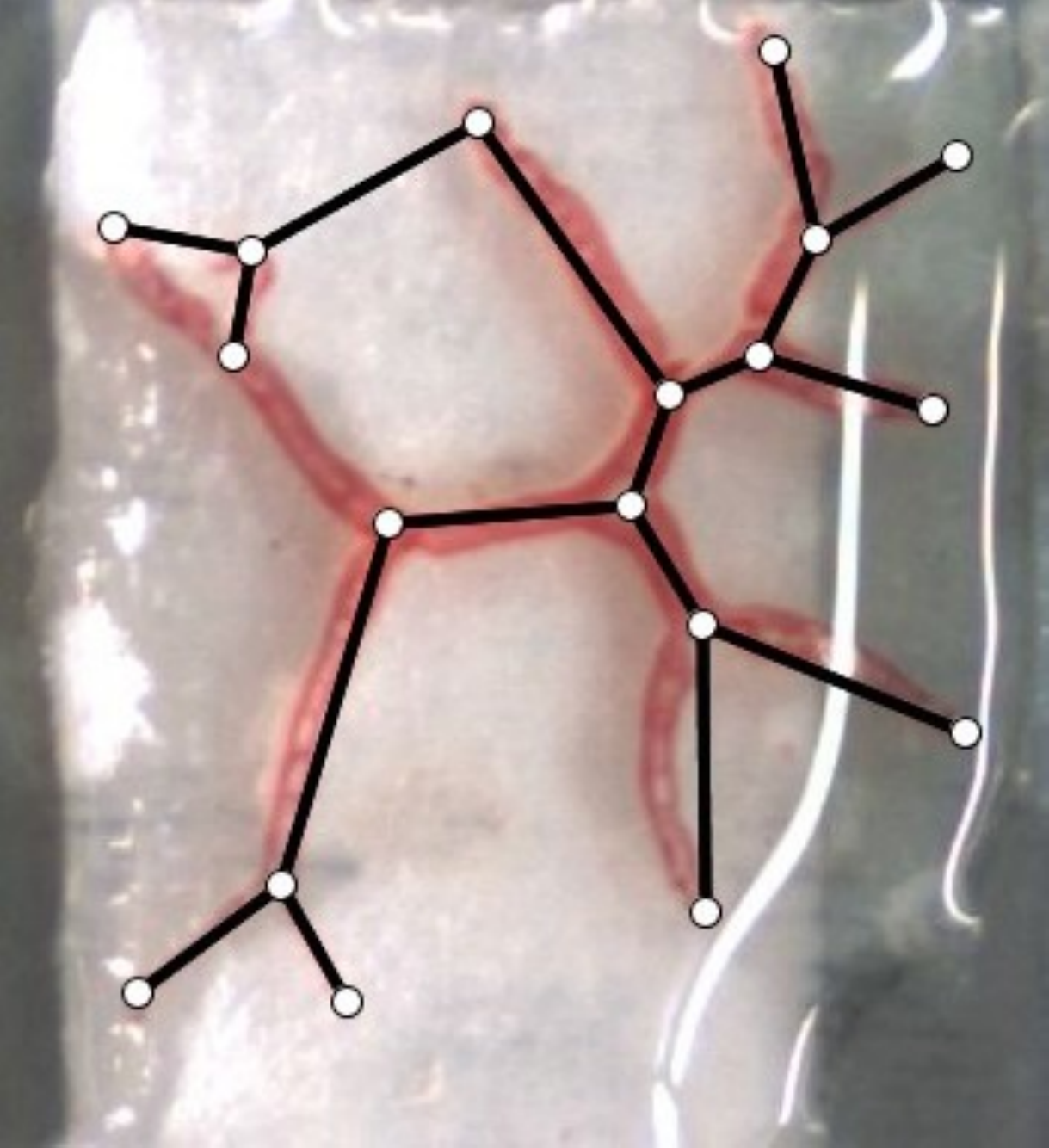}}
\subfigure[]{\includegraphics[scale=0.3]{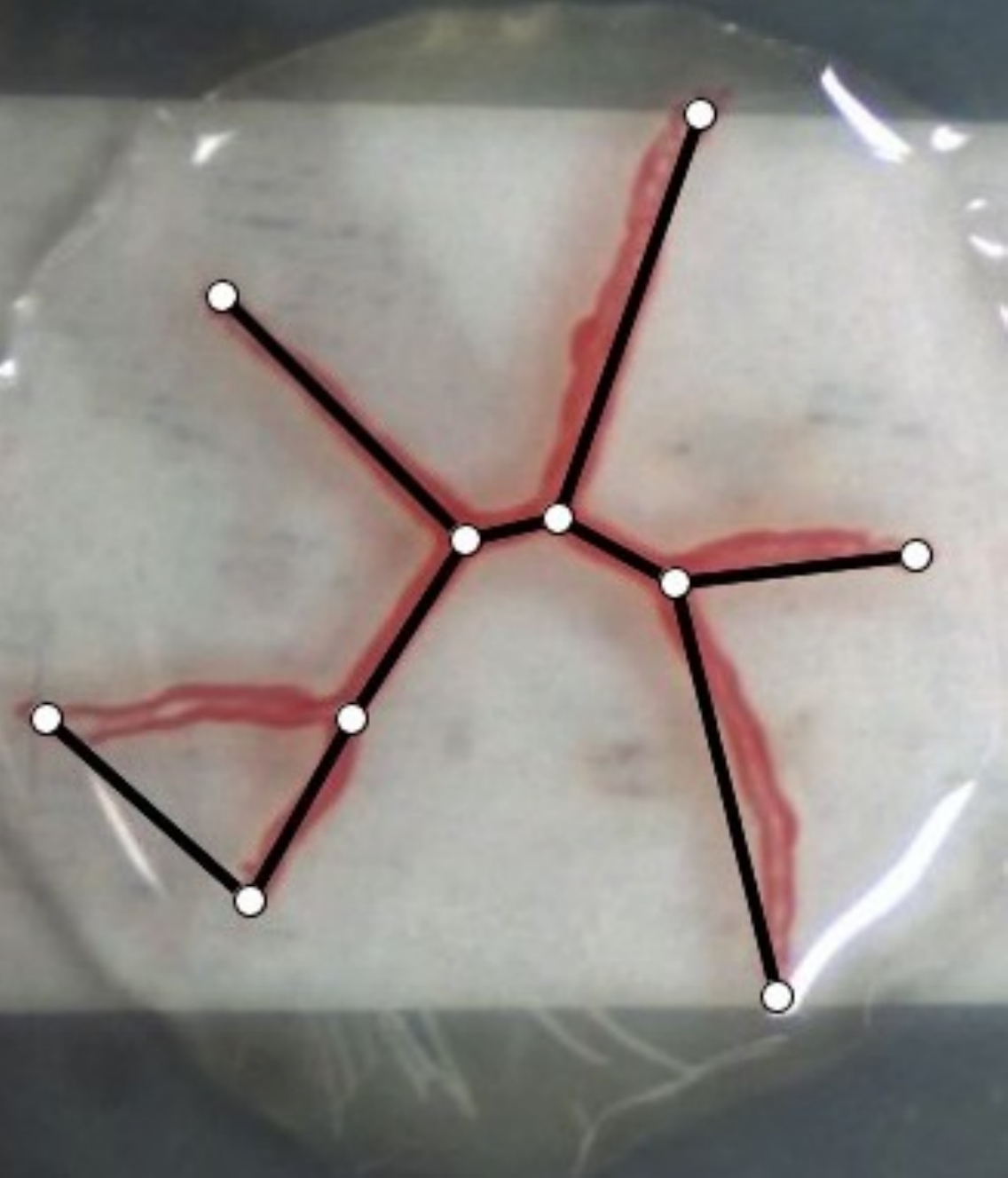}}
\caption{Approximating droplets with spanning trees. 
(a)~square control
(b)~round control,
(c)~square current,
(d)~round current
}
\label{fig:MST}
\end{figure}

The Euclidean minimum spanning tree (MST)~\cite{nevsetvril2001otakar} is a connected
acyclic graph which has minimum possible sum of edges’ lengths.

\begin{finding}
Skeletons of branching droplets are nearly minimum spanning trees. 
\end{finding}

The finding related to droplets which are single connected components. To verify if the statement is true we assigned branching sites as nodes of an Euclidean planar set and run our now classical~\cite{adamatzky1991neural} algorithm for constructing a minimum spanning tree grown from each of the nodes. In all of twenty branching droplets there was always a node from which such tree grown that it had a minimal sum of edge length and just one mismatching with the branching droplet. Example of minimum spanning trees constructed on branching sites of the droplets are shown in Fig.~\ref{fig:MST}.

\begin{figure}[!tbp]
    \centering
    \includegraphics[scale=0.3]{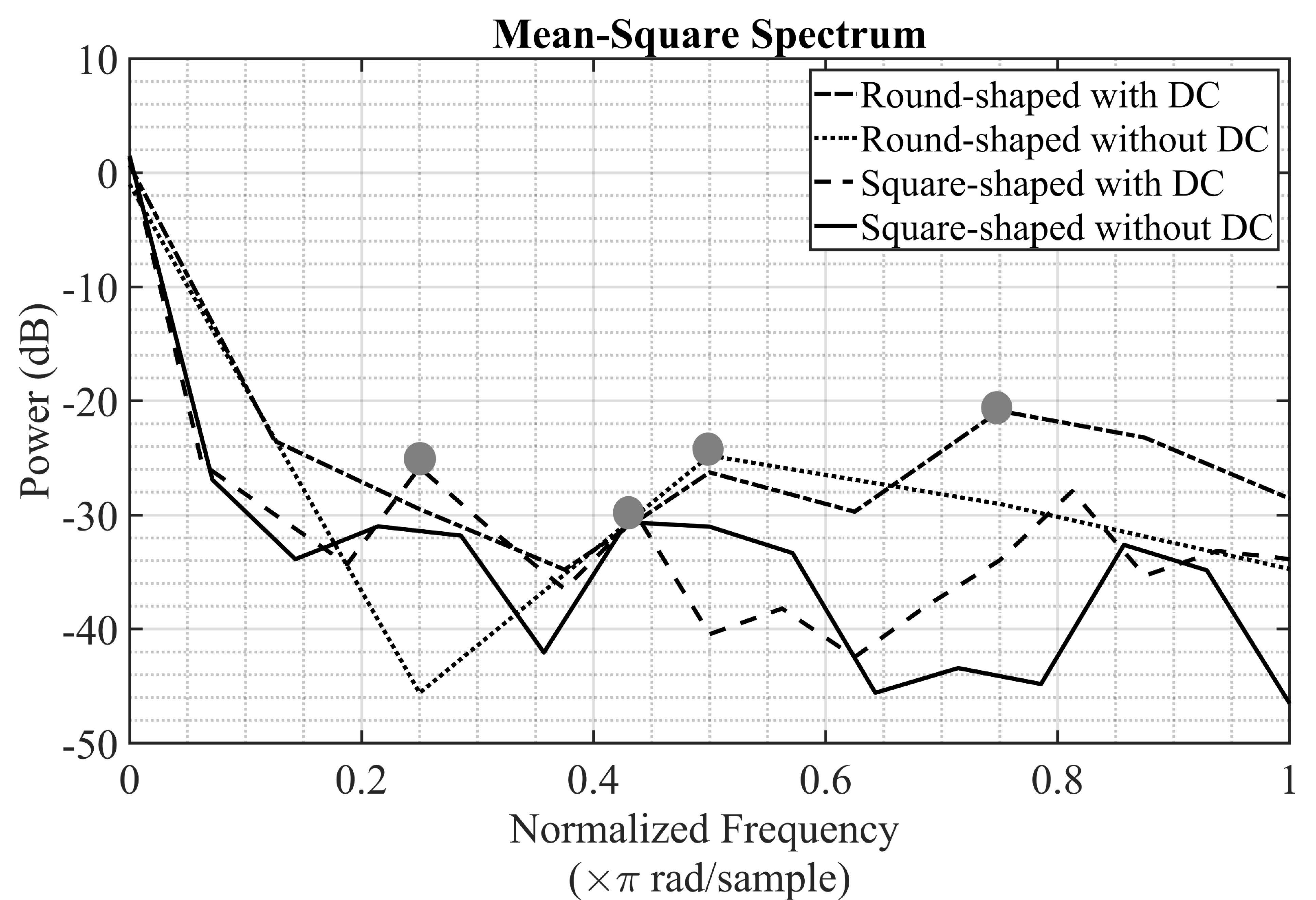}
    \caption{Dominating frequency of droplet's MR. Grey spots show the dominating frequency.}
    \label{fig:M3}
\end{figure}

\begin{finding}
A disorder of the branching droplets increases when DC is applied.
\end{finding}

\begin{figure}[!tbp]
    \centering
    \subfigure[]{\includegraphics[scale=0.25]{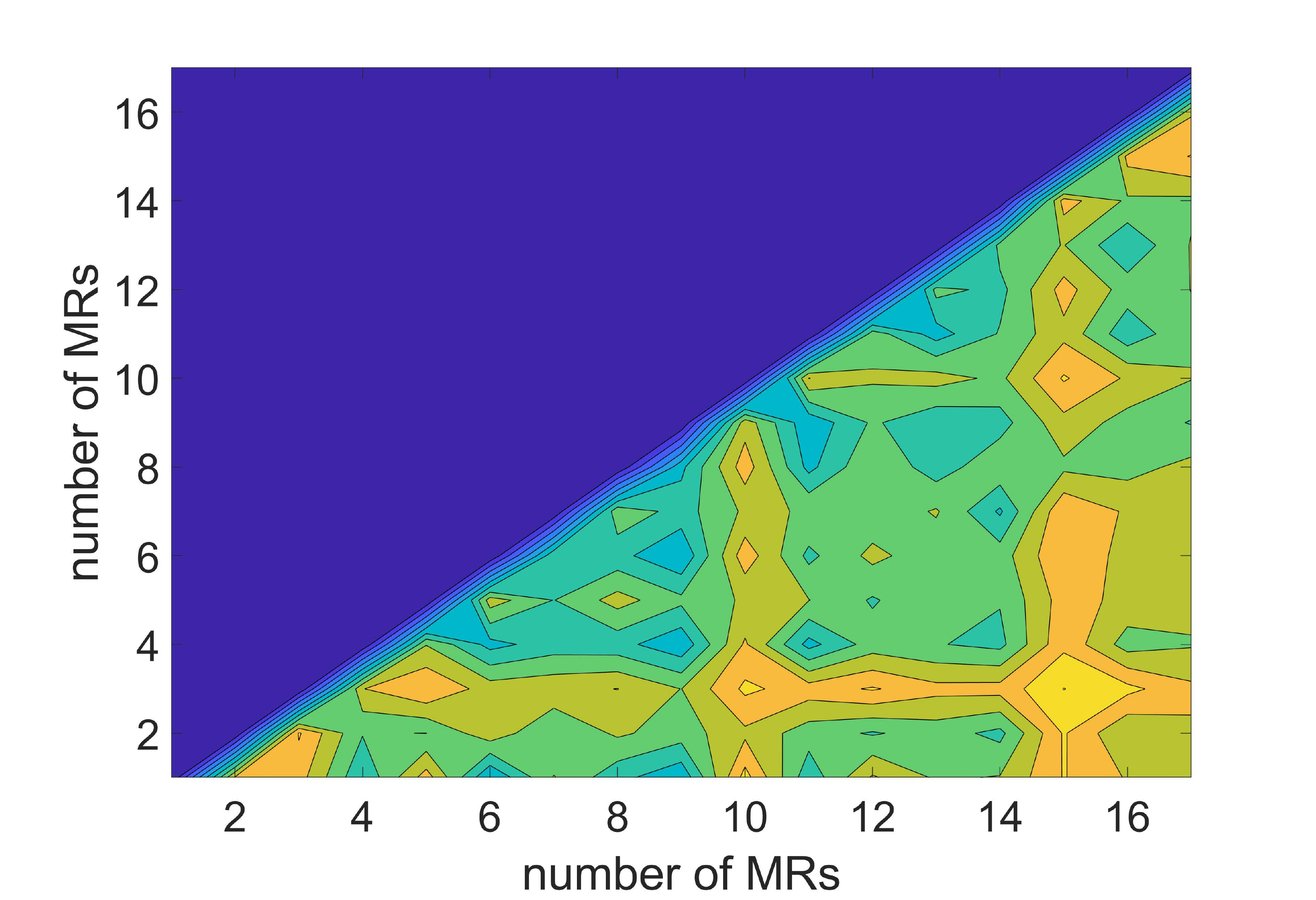}}
    \subfigure[]{\includegraphics[scale=0.25]{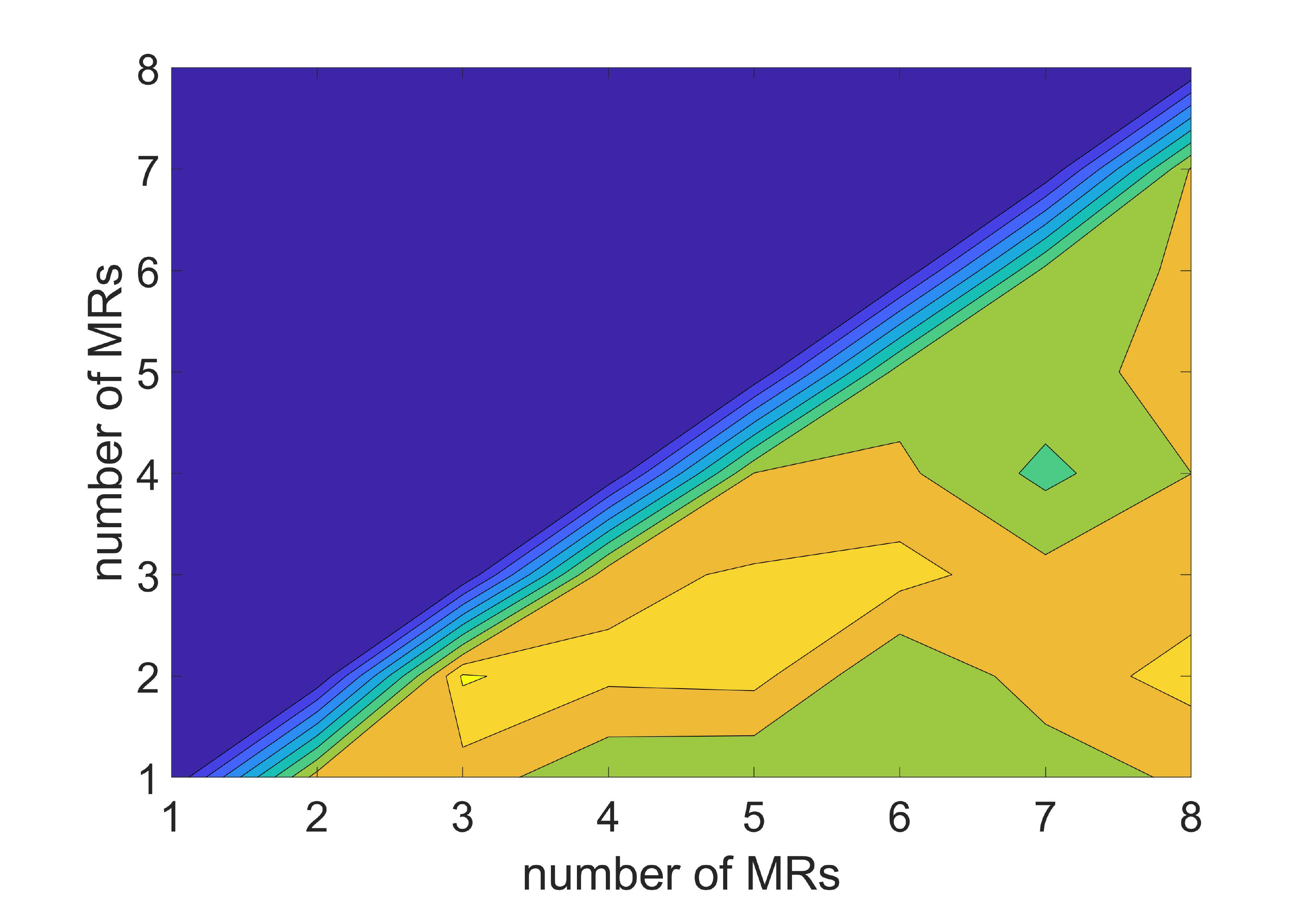}}
    \caption{The contour graph of distances between MRs (a) with an electrical field and (b) without an electrical field.}
    \label{fig:M4}
\end{figure}

Dominating frequencies (DF) of the entropy~\cite{adamatzky2019exploring} of MRs are shown in Fig.~\ref{fig:M3}. We can see that in the absence of DC DF for round-shaped droplets decreases while it increases for the square-shaped ones. Moreover, in the contour graph of distances (Fig.~\ref{fig:M4}) between MRs, one can see more nested patterns when an electrical field is applied. 

\begin{finding}
DC applied to the droplets increases their entropy.
\end{finding}

\begin{figure}[!tbp]
    \centering
    \subfigure[]{\includegraphics[scale=0.22]{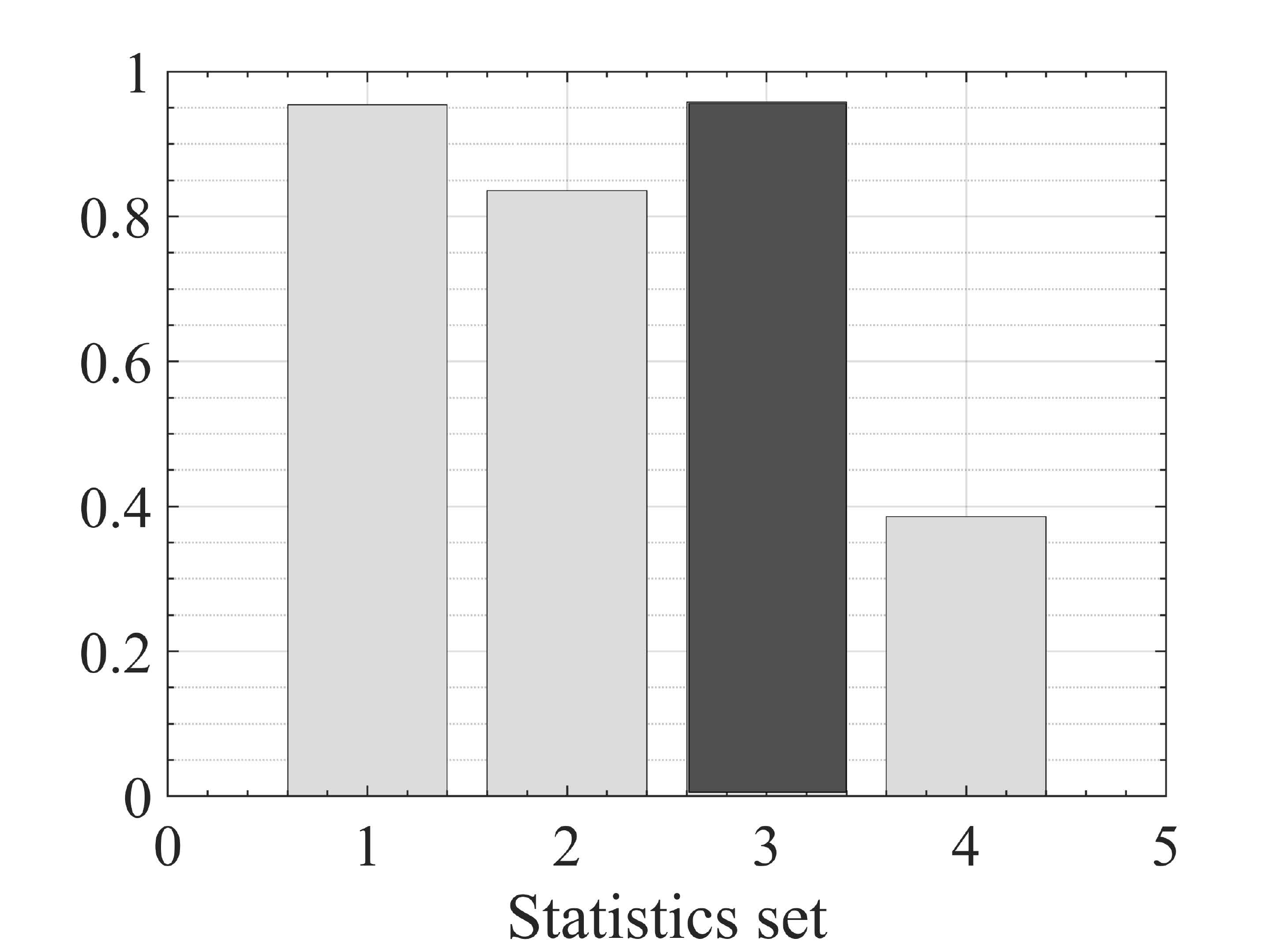}}
    \subfigure[]{\includegraphics[scale=0.22]{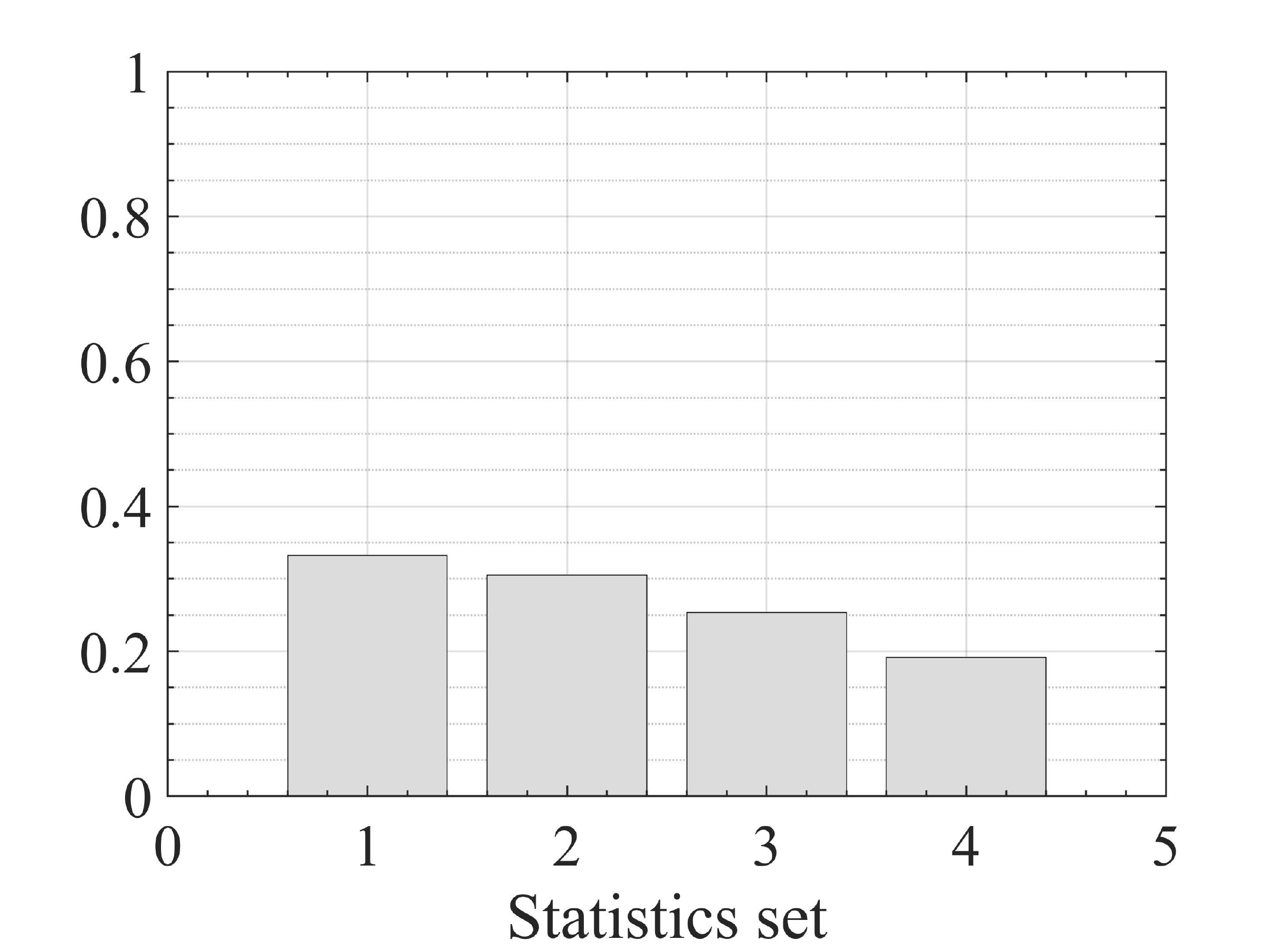}}
    \subfigure[]{\includegraphics[scale=0.22]{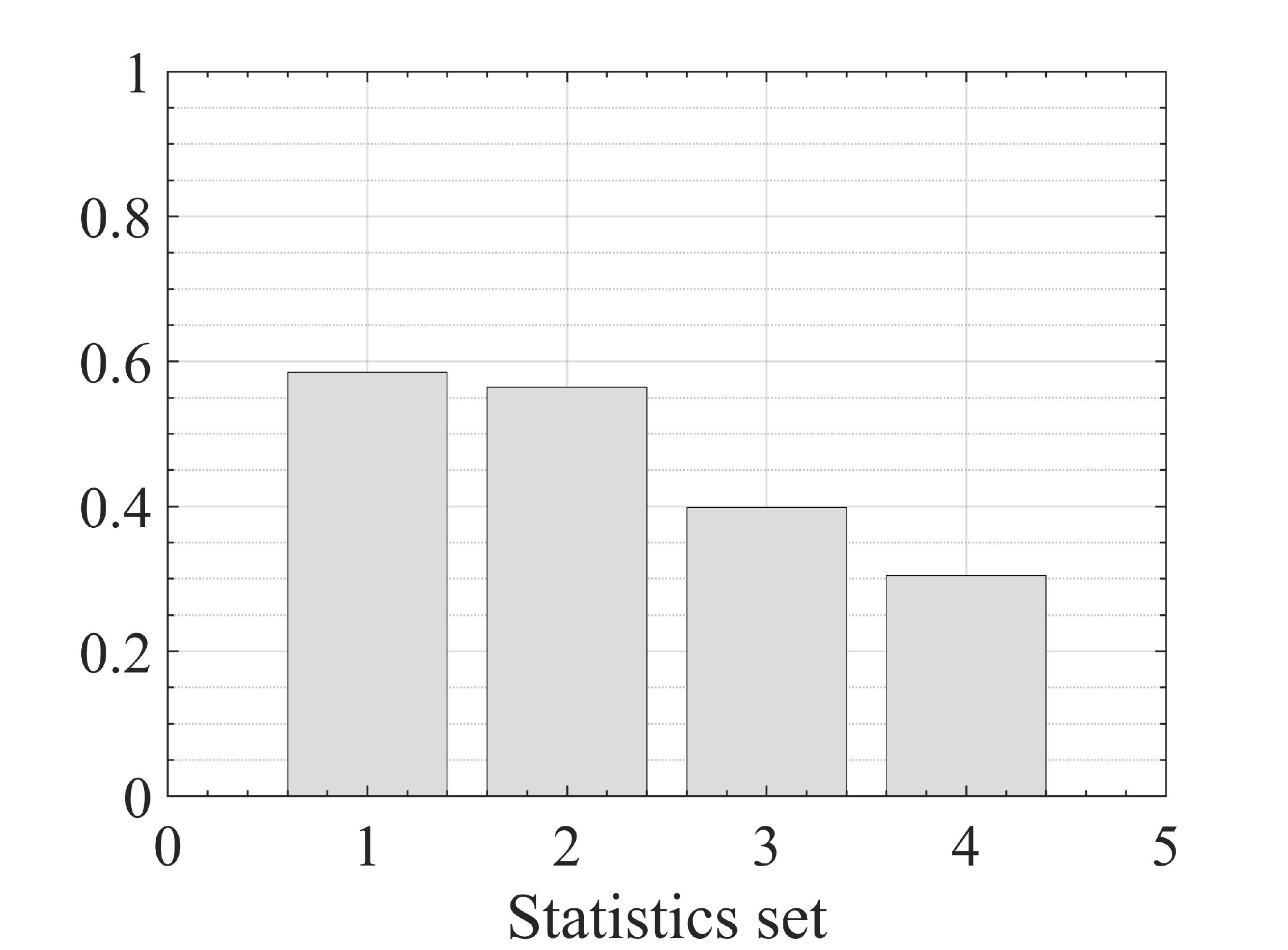}}
    \subfigure[]{\includegraphics[scale=0.22]{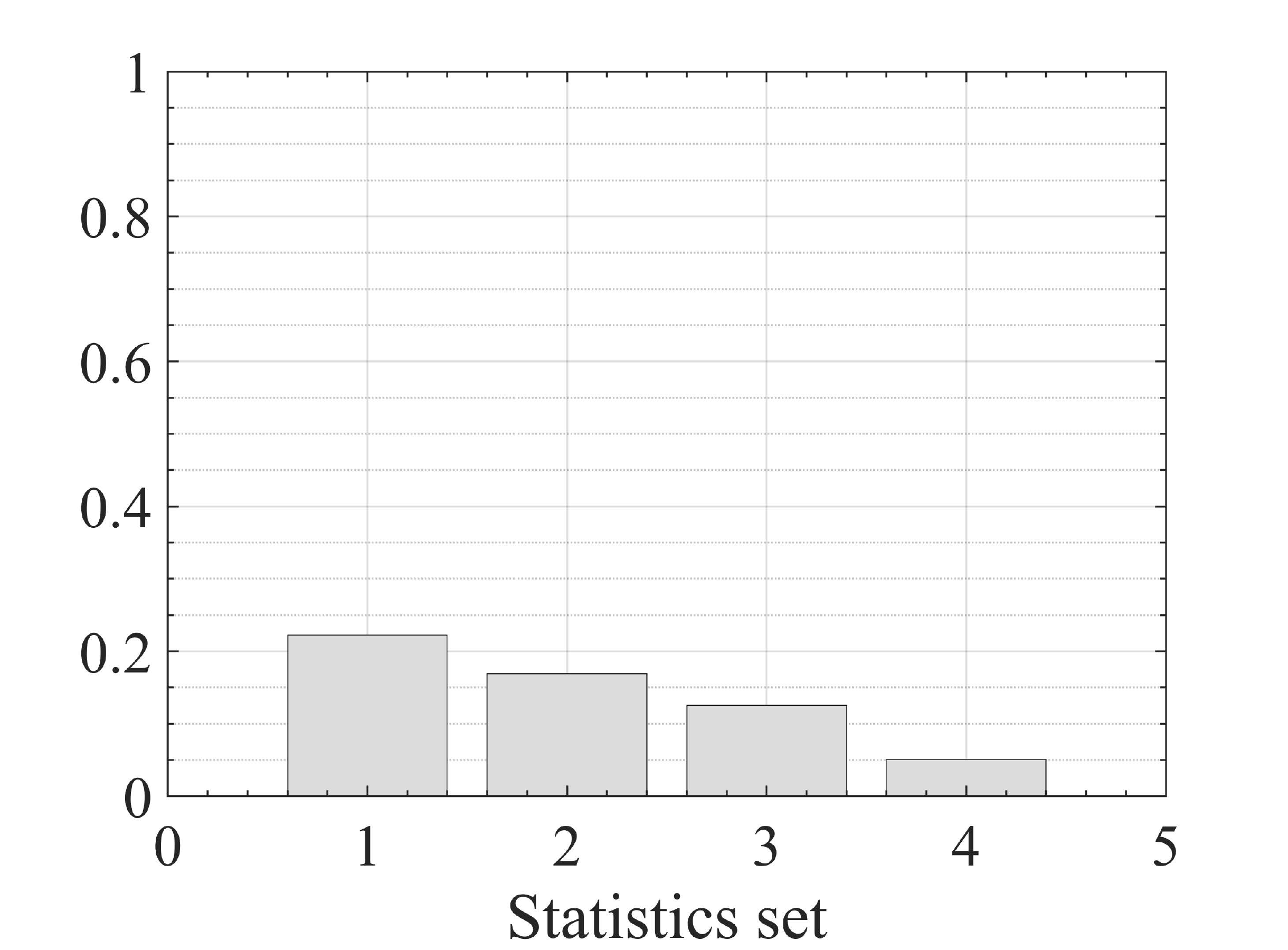}}
    \caption{Statistics for (a) $\mathbf{max}$, (b) $\mathbf{min}$, (c) $\mathbf{mean}$, and (d) $\mathbf{std}$. The $\mathbf{max}$ for square-shaped droplets subjected to DC is higher than the round-shaped ones in a similar condition. For that reason, the third column in (a) is shown in dark gray.}
    \label{fig:M5}
\end{figure}

When we calculated the entropy of the adjacency matrix of the created Delaunay triangulations, we summarise our findings in four sets of $St = \{\mathbf{max}, \mathbf{min}, \mathbf{mean}, \mathbf{std}\}$ where are respectively the maximum, minimum, average, and standard deviation of entropy related to instances belonging to $St$. $St_1$ and $St_2$ contain statistics of round-shaped droplets in the presence and absence of an electrical field, respectively. $St_3$ and $St_4$ have statistics of square-shaped droplets in the presence and absence of an electrical field, respectively. Figure~\ref{fig:M5} shows these statistics. One could find that, except for the $\mathbf{max}$, the entropy of droplets in square setup is lower than that of droplets in a round setup. The entropy of droplets subjected to DC is higher than the entropy of droplets without DC.

\section{Conclusion}

Direct current applied to decanol droplets in a thin layer of sodium decanoate with sodium chloride  does not affect directional growth of the branching droplets. The current increases complexity of the branching structures and their entropy.  Further studies are required to uncover chemical and physical mechanisms of the reported phenomena.

\section*{Acknowledgement}

J.\v{C}. and D.\v{S} were supported by the Czech Science Foundation (17-21696Y). 

\bibliographystyle{plain}
\bibliography{manuscript.bbl}

\begin{thebibliography}{10}

\bibitem{adamatzky1991neural}
A~Adamatzky.
\newblock Neural algorithm for constructing minimal spanning tree.
\newblock {\em Neural Network World}, 6:335--339, 1991.

\bibitem{adamatzky2019exploring}
A.~Adamatzky and M.~M. Dehshibi.
\newblock Exploring tehran with excitable medium.
\newblock In A.~Adamatzky, S.~Akl, and G.~Sirakoulis, editors, {\em From
  Parallel to Emergent Computing}, pages 475--488. CRC Press, 2019.

\bibitem{adamatzky2000choosey}
Andrew Adamatzky.
\newblock Choosey hot sand: reflection of grain sensitivity on pattern
  morphology.
\newblock {\em International Journal of Modern Physics C}, 11(01):47--68, 2000.

\bibitem{adamatzky2005reaction}
Andrew Adamatzky, Benjamin De~Lacy Costello, and Tetsuya Asai.
\newblock {\em Reaction-diffusion computers}.
\newblock Elsevier, 2005.

\bibitem{adamatzky2010generative}
Andrew Adamatzky and Genaro~J Martinez.
\newblock On generative morphological diversity of elementary cellular
  automata.
\newblock {\em Kybernetes}, 39(1):72--82, 2010.

\bibitem{biasotti2000extended}
Silvia Biasotti, Bianca Falcidieno, and Michela Spagnuolo.
\newblock Extended reeb graphs for surface understanding and description.
\newblock In {\em International conference on discrete geometry for computer
  imagery}, pages 185--197. Springer, 2000.

\bibitem{vcejkova2018multi}
Jitka {\v{C}}ejkov{\'a}, Martin~M Hanczyc, and Franti{\v{s}}ek
  {\v{S}}t{\v{e}}p{\'a}nek.
\newblock Multi-armed droplets as shape-changing protocells.
\newblock {\em Artificial life}, 24(1):71--79, 2018.

\bibitem{cejkova2016evaporation}
Jitka {\v{C}}ejkov{\'a}, Franti{\v{s}}ek {\v{S}}t{\v{e}}p{\'a}nek, and
  Martin~M. Hanczyc.
\newblock Evaporation-induced pattern formation of decanol droplets.
\newblock {\em Langmuir}, 32(19):4800--4805, 2016.

\bibitem{dehshibi2017hybrid}
Mohammad~Mahdi Dehshibi, Mohamad Sourizaei, Mahmood Fazlali, Omid Talaee,
  Hossein Samadyar, and Jamshid Shanbehzadeh.
\newblock A hybrid bio-inspired learning algorithm for image segmentation using
  multilevel thresholding.
\newblock {\em Multimedia Tools and Applications}, 76(14):15951--15986, 2017.

\bibitem{gholami2018complexity}
Neda Gholami, Mohammad~Mahdi Dehshibi, Mahmood Fazlali, Antonio Rueda-Toicen,
  Hector Zenil, and Andrew Adamatzky.
\newblock On complexity of post-processing in analyzing gate-driven x-ray
  spectrum.
\newblock {\em arXiv preprint arXiv:1807.09063}, 2018.

\bibitem{lazarus1999level}
Francis Lazarus and Anne Verroust.
\newblock Level set diagrams of polyhedral objects.
\newblock In {\em Proceedings of the fifth ACM symposium on Solid modeling and
  applications}, pages 130--140. ACM, 1999.

\bibitem{lempel1976complexity}
Abraham Lempel and Jacob Ziv.
\newblock On the complexity of finite sequences.
\newblock {\em IEEE Transactions on information theory}, 22(1):75--81, 1976.

\bibitem{nevsetvril2001otakar}
Jaroslav Ne{\v{s}}et{\v{r}}il, Eva Milkov{\'a}, and Helena
  Ne{\v{s}}et{\v{r}}ilov{\'a}.
\newblock Otakar {B}o\r{u}vka on minimum spanning tree problem translation of
  both the 1926 papers, comments, history.
\newblock {\em Discrete mathematics}, 233(1-3):3--36, 2001.

\bibitem{ramin2012counting}
Marjan Ramin, Alireza Sepas-Moghaddam, Payam Ahmadvand, and Mohammad~Mahdi
  Dehshibi.
\newblock Counting the number of cells in immunocytochemical images using
  genetic algorithm.
\newblock In {\em Hybrid Intelligent Systems (HIS), 2012 12th International
  Conference on}, pages 185--190. IEEE, 2012.

\bibitem{sethian1999level}
James~Albert Sethian.
\newblock {\em Level set methods and fast marching methods: evolving interfaces
  in computational geometry, fluid mechanics, computer vision, and materials
  science}, volume~3.
\newblock Cambridge university press, 1999.

\bibitem{taghipour2016complexity}
Nassim Taghipour, Hamid Haj~Seyyed Javadi, Mohammad~Mahdi Dehshibi, and Andrew
  Adamatzky.
\newblock On complexity of persian orthography: L-systems approach.
\newblock {\em Complex Systems}, 25(2):127--156, 2016.

\end{thebibliography}

\end{document}